\documentclass[hidelinks]{article}

\PassOptionsToPackage{quiet}{fontspec} 

\usepackage{algorithm}
\usepackage{algorithmic}
\usepackage{amsmath}
\usepackage{amssymb}
\usepackage{amsthm}
\usepackage{anyfontsize}
\usepackage{abstract}
\usepackage{booktabs}
\usepackage{enumerate}
\usepackage{enumitem}
\usepackage{float}
\usepackage{times}
\usepackage{geometry}
\usepackage{graphicx}
\usepackage{lmodern}
\usepackage{indentfirst}
\usepackage{mathrsfs}
\usepackage{multirow}
\usepackage{ragged2e}
\usepackage[figuresright]{rotating}
\usepackage{setspace}
\usepackage{subcaption}
\usepackage{threeparttable}
\usepackage{url}
\usepackage{xcolor}
\usepackage{xr}

\usepackage[T1]{fontenc}
\usepackage[utf8]{inputenc}
\usepackage{lmodern}
\usepackage[english]{babel}
\usepackage[autostyle]{csquotes}

\usepackage[labelfont=bf, labelsep=period, font=footnotesize]{caption}

\def\wh{\widehat}

\def\ba{\boldsymbol\alpha}
\def\bb{\boldsymbol\beta}
\def\bC{\boldsymbol{C}}
\def\bE{\boldsymbol{E}}
\def\bF{\boldsymbol{F}}
\def\bH{\boldsymbol{H}}
\def\bI{\boldsymbol{I}}
\def\bM{\boldsymbol{M}}
\def\bP{\boldsymbol{P}}
\def\bR{\boldsymbol{R}}
\def\bx{\boldsymbol{x}}
\def\bX{\boldsymbol{X}}
\def\bV{\boldsymbol{V}}
\def\mE{\mathbb{E}}
\def\mF{\mathcal{F}}
\def\mO{\mathcal{O}}

\usepackage{natbib}
\setcitestyle{open = {(}, close = {)}}
\usepackage{hyperref}
\definecolor{citecolor}{HTML}{7F0000}
\hypersetup{
    colorlinks = true,
    urlcolor = black,
    linkcolor = citecolor,
    anchorcolor = citecolor,
    citecolor = citecolor
}

\newtheorem{assumption}{Assumption}
\newtheorem{proposition}{Proposition}
\newtheorem{theorem}{Theorem}
\newtheorem{remark}{Remark}

\begin{document}

\title{High-Dimensional Matrix-Variate Diffusion Index Models for Time Series Forecasting\thanks{This work is supported in part by the National Natural Science Foundation of China (NSFC) under grant numbers 12201558 (Z.G.) and 12371272 (R. Zhang).}}

\date{}


\author{
    Zhiren Ma\textsuperscript{1}
    \and 
    Qian Zhao\textsuperscript{1,}\thanks{Co-first author.}
    \and 
    Riquan Zhang\textsuperscript{1}
    \and 
    Zhaoxing Gao\textsuperscript{2,}\thanks{Corresponding author.}
}
\footnotetext[1]{\scriptsize School of Statistics and Data Science, Shanghai University of International Business and Economics. \\
Email: \url{mazhirenshnu@gmail.com} (Z. Ma), \url{zhaoqian@suibe.edu.cn} (Q. Zhao), \url{zhangriquan@163.com} (R. Zhang).}
\footnotetext[2]{\scriptsize School of Mathematical Sciences, University of Electronic Science and Technology of China. \\
Email: \url{zhaoxing.gao@uestc.edu.cn}.}

\maketitle


\renewcommand{\abstractname}{Abstract} 
\begin{abstract}
\begin{spacing}{1.5}
This paper proposes a novel diffusion-index model for forecasting when predictors are high-dimensional matrix-valued time series. We apply an $\alpha$-PCA method to extract low-dimensional matrix factors and build a bilinear regression linking future outcomes to these factors, estimated via iterative least squares. To handle weak factor structures, we introduce a supervised screening step to select informative rows and columns. Theoretical properties, including consistency and asymptotic normality, are established. Simulations and real data show that our method significantly improves forecast accuracy, with the screening procedure providing additional gains over standard benchmarks in out-of-sample mean squared forecast error.
\bigskip{} \par
\end{spacing}
\end{abstract}

\newpage
\pagenumbering{arabic}
\setstretch{1.5}


\section{Introduction}\label{sec1}
\justifying
\setstretch{1.5}
Time series forecasting has a wide range of applications across diverse fields, including biology, medicine, geography, economics, and financial markets. Extensive research has been conducted on both univariate (\cite{box1976time, engle1982autoregressive, engle1995multivariate, tong1990non, brockwell1991time, tiao1981modeling, tsay2005analysis}) and multivariate (\cite{tiao1989model,stock2005empirical, lutkepohl2005new, tsay2013multivariate}) time series, resulting in the development of a variety of classical and foundational models such as ARIMA (\cite{box1970time}), GARCH (\cite{bollerslev1986generalized}), and VAR (\cite{sims1982policy}). Another well-established forecasting framework that has received considerable attention is the diffusion-index or factor-augmented method proposed by \cite{Stock01122002, Stock01042002}. This methodology is particularly effective in scenarios characterized by a large number of predictors relative to the number of target variables, which is a common occurrence in modern data-rich environments. With the advancement of computational power and statistical methodology, there has been increasing scholarly interest in modeling high-dimensional time series (\cite{buhlmann2011statistics, lam2011estimation, lam2012factor, belloni2013least, javanmard2014confidence, chang2015high, adamek2023lasso}). In recent years, the prevalence of matrix-variate data in big data contexts has further challenged traditional modeling paradigms. In particular, the conventional vectorization of matrices often disrupts the intrinsic structural dependencies within the data, leading to potential information loss. Consequently, there is a pressing need to develop new statistical methodologies that can directly model matrix-variate time series while preserving their inherent structural properties.

Numerous studies have investigated matrix-variate data, including works by \cite{fan2011high, fan2013large, leng2012sparse, yin2012model, zhao2014structured, zhou2014regularized, wang2016efficient, gao2020segmenting, chang2023modelling, han2024simultaneous}, among others. However, none of these can be used to forecast another specific scalar target. For instance, a special case of the Matrix Autoregression (MAR) model, as proposed by \cite{chen2021autoregressive}, is considered to predict an economic index, i.e. the GDP growth of major OECD countries, denoted as \(y_{t+h}\), using matrix predictors \(\bX_t\). This relationship can be formalized in the following Equation (\ref{eq-raw}):
\begin{equation}
y_{t+h} = \ba_1^{\prime} \bX_t \bb_1 + e_{t+h},
\label{eq-raw}
\end{equation}
where \(y_{t+h}\) denotes the scalar response at forecast horizon \(h\), \(\bX_t\) represents the matrix-valued predictors at time \(t\), and \(\ba_1\) and \(\bb_1\) are the row and column regression coefficients, respectively. As the dimensions of the data increases, incorporating a large number of predictors can lead to overparameterization and multicollinearity, thereby degrading forecasting performance and making the aforementioned model unsuitable for such high-dimensional settings. This highlights the necessity for effective dimension reduction techniques to address the ``curse of dimensionality" and reduce the number of parameters to be estimated. A wide body of literature has addressed this issue, most notably through Principal Component Analysis (PCA) (\cite{anderson1958introduction, anderson1963asymptotic, stock2005empirical, gu2020empirical, he2023shrinking, huang2022scaled, chen2023statistical, gao2019structural, gao2023divide, gao2023two, gao2025supervised}) and a variety of factor-based models (\cite{bai2003inferential, bai2002determining, bai2006confidence, bai2008forecasting, bai2023approximate, lam2011estimation, lam2012factor, ahn2013eigenvalue, chang2015high, bernanke2005measuring, liu2016regime, wang2019factor}). By leveraging factor methods, the challenges inherent in high-dimensional forecasting can be alleviated, estimation accuracy can be enhanced, and the predictor space can be efficiently compressed by capturing underlying low-dimensional latent structures. Nevertheless, existing research on PCA and factor models has primarily focused on contemporaneous relationships, while their application to dynamic time series forecasting has received comparatively less attention. Furthermore, current research on diffusion index models predominantly focuses on vector-based representations, whereas extensions to matrix-variate structure remain relatively underdeveloped.

In this paper, we adopt the factor-based dimensionality reduction methodology proposed by \cite{chen2023statistical}, known as \textit{$\alpha$-PCA}, to extract the latent factor matrix. Additionally, we employ the iterative least-squares estimation (\textit{LSE}) procedure introduced by \cite{chen2021autoregressive} to estimate the row and column loading vectors. The \textit{$\alpha$-PCA} method overcomes the limitations of conventional 2D-PCA, which relies solely on variance-based analysis. Instead, it incorporates both the first and second moments of the observed matrix through a spectral aggregation process, thereby balancing data informativeness with the parameter $\alpha$ and capturing more information than traditional PCA. The \textit{LSE} method adopts an iterative-convergence strategy, where one estimator in the pair is fixed while the other is updated until the convergence criteria are satisfied. Our estimation procedure involves two consecutive stages. First, given the observed matrix, we apply \textit{$\alpha$-PCA} to estimate the row and column loading matrices and recover the latent factor process. Second, using the latent factor matrix obtained in the first stage and the univariate time series of indices, we estimate the left and right loading vectors via the \textit{LSE} method and perform forecasting analysis. Moreover, we propose a novel supervised screening refinement step that incorporates the correlation between the target index and the observation matrix. Specifically, we eliminate rows and columns of the observation matrix that exhibit low correlation with the target index, resulting in a refined observation matrix. This refined matrix is then used within the same estimation and forecasting framework. We find that this refinement step effectively reduces noise in the raw matrix and enhances the accuracy of our final predictions. A detailed description of the model is provided in Section \ref{sec2}, along with the refinement procedure in Section \ref{sec4}.

The structure of this paper is organized as follows. Section \ref{sec2} outlines the construction of the main model and the associated estimation methodology. Section \ref{sec3} examines the theoretical properties of the latent factor matrix and the loading vectors. In Section \ref{sec4}, we introduce a novel supervised screening refinement step for index forecasting. Section \ref{sec5} presents the simulation results, assessing the finite sample performance of our proposed model. Section \ref{sec6} provides an empirical analysis based on real-world data. Finally, Section \ref{sec7} concludes the paper and offers suggestions for future research. Proofs of theoretical results and additional materials for the simulations and empirical studies, including data descriptions, are provided in the online Appendix.

\subsection{Notations}
\justifying
\setstretch{1.5}

We use the following notations. Let $x$, $\bx$, and $\bX$ represent a scalar, a vector, and a matrix, respectively. The $i$-th row, $j$-th column, and $(i,j)$-th element of a matrix $\bX$ are denoted as $\bX_{i\cdot}$, $\bX_{\cdot j}$, and $x_{ij}$, respectively. For a vector $\bx = (x_1, \ldots, x_p)^{\prime}$ of dimension $p \times 1$, the $\ell_1$-norm is defined as $\displaystyle \Vert \bx \Vert_1 = \sum_{i=1}^{p} |x_i|$, and the $\ell_2$-norm is given by $\displaystyle \Vert \bx \Vert_2 = \sqrt{\sum_{i=1}^{p} x_i^2}$. For a matrix $\bH$, the $\ell_1$-norm is defined as $\displaystyle \Vert \bH \Vert_1 = \max_j \sum_i |x_{ij}|$, the $\ell_2$-norm is defined as $\displaystyle \Vert \bH \Vert_2 = \sqrt{\lambda_{\mathrm{max}}(\bH^{\prime} \bH)}$, where $\lambda_{\mathrm{max}}(\cdot)$ denotes the largest eigenvalue of a matrix, and the Frobenius norm is defined as $\displaystyle \Vert \bH \Vert_F = \sqrt{\mathrm{tr}(\bH\bH^{\prime})}$, where $\mathrm{tr}(\cdot)$ denotes the trace of a matrix. The superscript $^\prime$ denotes the transpose of a vector or matrix. The symbol $\otimes$ represents the Kronecker product, and $\textrm{vec}(\cdot)$ denotes the vectorized form of a matrix. Additionally, we use the notation $a \asymp b$ to indicate that $a = O(b)$ and $b = O(a)$. Finally, $C$ represents a universal constant, which may vary throughout the paper and across different lines.

\section{Model and Methodology}\label{sec2}
\justifying
\setstretch{1.5}

Let \(\{\bX_1, \ldots, \bX_T\}\) be a sequence of matrix-valued time series, where \(\bX_t \in \mathbb{R}^{m \times n}\) for \(1 \leq t \leq T\). We assume that both \(m\) and \(n\) are large, implying that \(\bX_t\) represents high-dimensional random sequences. Let \(\{y_1, \ldots, y_T\}\) denote a sequence of scalar time series. Our goal is to predict \(y_{t+h}\) using \(\bX_t\). Since \(\bX_t\) is large, it is difficult to apply the bi-linear regression approach discussed in Section \ref{sec1}. To address this issue, we propose a matrix-variate diffusion index model, which is formulated as follows:
\begin{equation}
\left\{
    \begin{array}{ll}
        \bX_t = \bR \bF_t \bC^{\prime} + \bE_t, \\
        y_{t+h} = \ba^{\prime} \bF_t \bb + e_{t+h}, \\
    \end{array}
\right.
\label{eq-model}
\end{equation}
where \(\bF_t \in \mathbb{R}^{k \times r}\) denotes the latent factor matrix (or matrix diffusion index), \(\bR \in \mathbb{R}^{p \times k}\), \(\bC \in \mathbb{R}^{q \times r}\), \(\ba \in \mathbb{R}^{k \times 1}\), and \(\bb \in \mathbb{R}^{r \times 1}\) are the factor loading matrices and vectors that need to be estimated. $\bE_t$ represents the noise matrix, while $e_{t+h}$ denotes the innovations at time \(t+h\).

Note that the latent factor matrix $\bF_t$, the loading matrices $\bR$ and $\bC$, along with the loading vectors $\ba$ and $\bb$, are not identifiable. Specifically, the triplets $(\bR, \bF_t, \bC^{\prime})$ and $(\ba^{\prime}, \bF_t, \bb)$ can be replaced by $(\bR \bH_R, \bH_R^{-1} \bF_t \bH_C^{-1 \prime}, \bH_C^{\prime} \bC^{\prime})$ and $(\ba^{\prime} \bH_R, \bH_R^{-1} \bF_t \bH_C^{-1 \prime}, \bH_C^{\prime} \bb)$ respectively, without altering the original model structure, where \(\bH_R \in \mathbb{R}^{k \times k}\) and \(\bH_C \in \mathbb{R}^{r \times r}\) denote the corresponding rotation matrices. Consequently, instead of estimating the ground truth components $(\bR, \bF_t, \bC, \ba, \bb)$ as discussed in \cite{chen2023statistical}, our goal is to estimate the rotational form, or equivalently, the column space of these components. To achieve this, we impose some constraints on the loading matrices and vectors such that
\[
\frac{1}{p} \bR^{\prime} \bR = \bI_k \ \ \text{and} \ \ \frac{1}{q} \bC^{\prime} \bC = \bI_r,
\]
where $\bI_n$ denotes the $n \times n$ identity matrix. For the estimation of \(\ba\) and \(\bb\), which rely on the estimation of \(\bF_t\), it is evident from the theoretical properties that there exists a matrix \(\bH_R\) such that the relationship holds consistently. Therefore, we assume that \(\Vert \bH_R^{\prime} \ba \Vert_2 = 1\), which ensures the identification conditions for both \(\ba\) and \(\bb\). Further details on this are deferred to Section \ref{sec2.2}.

Based on the framework above, we first employ the \textit{$\alpha$-PCA} method proposed by \cite{chen2023statistical} to estimate the loading matrices and the latent factor matrix. Next, we apply the \textit{LSE} procedure introduced by \cite{chen2021autoregressive} to estimate the corresponding loading vectors. The complete estimation procedure is outlined as follows.

\subsection{Estimation of Latent Factor Matrix}
\justifying
\setstretch{1.5}

Consider a sequence of matrix-valued time series observations $\{\bX_1, \ldots, \bX_T\}$, where $\bX_t \in \mathbb{R}^{p \times q}$ for $1 \leq t \leq T$, and corresponding scalar time series responses $\{y_1, \ldots, y_T\}$, with $y_t \in \mathbb{R}$ for $1 \leq t \leq T$. To address the forecasting task, we begin with estimating the latent factor matrix using the \textit{$\alpha$-PCA} method. In the context of the model in Equation (\ref{eq-model}), this method aggregates information from both the first and second moments of the observed matrices, extracting the latent factors via a spectral decomposition approach. Specifically, we define the following statistics:
\begin{equation}
\widehat{\bM}_R := \frac{1}{pq} \left( (1+\alpha) \cdot \overline{\bX} \overline{\bX}^\prime + \frac{1}{T} \sum_{t=1}^T ( \bX_t - \overline{\bX} ) ( \bX_t - \overline{\bX} )^\prime \right),
\label{eq-MR}
\end{equation}
\begin{equation}
\widehat{\bM}_C := \frac{1}{pq} \left( (1+\alpha) \cdot \overline{\bX}^\prime \overline{\bX} + \frac{1}{T} \sum_{t=1}^T ( \bX_t - \overline{\bX} )^\prime ( \bX_t - \overline{\bX} ) \right).
\label{eq-MC}
\end{equation}
Here, $\alpha \in [-1, \infty)$ is a hyperparameter that balances the contributions of the first and second moments. The sample mean is denoted by $\displaystyle\overline{\bX} = \frac{1}{T} \sum_{t=1}^{T} \bX_t$. The matrices \(\displaystyle\frac{1}{T} \sum_{t=1}^T ( \bX_t - \overline{\bX} ) ( \bX_t - \overline{\bX} )^\prime\) and \(\displaystyle\frac{1}{T} \sum_{t=1}^T ( \bX_t - \overline{\bX} )^\prime ( \bX_t - \overline{\bX} )\) represent the sample row and column covariance matrices, respectively. The estimators $\wh{\bR}$ and $\wh{\bC}$ are then obtained as $\sqrt{p}$ times the top $k$ eigenvectors of $\widehat{\bM}_R$, and $\sqrt{q}$ times the top $r$ eigenvectors of $\widehat{\bM}_C$, ordered by descending eigenvalues. Here, $k$ and $r$ represent the row and column dimensions of the latent factor matrix $\bF_t$, respectively. Without loss of generality, we further impose the constraints that $\displaystyle\frac{1}{p} \wh{\bR}^\prime \wh{\bR} = \bI_p$ and $\displaystyle\frac{1}{q} \wh{\bC}^\prime \wh{\bC} = \bI_q$ on the estimators.

Then, by multiplying both sides of the equation $\bX_t = \wh{\bR} \wh{\bF}_t \wh{\bC}^\prime$ with the inverse of loading matrices $\wh{\bR}$ and $\wh{\bC}^\prime$, we obtain the estimated latent factor matrix $\wh{\bF}_t$ as
\[
\wh{\bF}_t = \frac{1}{pq} \wh{\bR}^\prime \bX_t \wh{\bC},
\]
where $\wh{\bF}_t$ is consistent with the rotated factor matrix $\bH_R^{-1} \bF_t \bH_C^{\prime -1}$, with $\bH_R$ and $\bH_C$ representing the row and column rotation matrices, respectively, as defined in Equations (12) and (13) of \cite{chen2023statistical}.

To simplify the notation, we introduce the following transformed variables.

Let $\displaystyle\widetilde\alpha = \sqrt{\alpha+1} - 1$ and 
\[\bX_t^{\star} := \bX_t + \widetilde\alpha\overline{\bX}, \ \widetilde\bF_t := \bF_t + \widetilde\alpha\overline{\bF} \ \text{and} \ \widetilde\bE_t := \bE_t + \widetilde\alpha\overline{\bE},\]
then we have
\[\displaystyle\bX_t^{\star} = \bR \widetilde\bF_t \bC^{\prime} + \widetilde{\bE}_t,\]
which means Equations (\ref{eq-MR}) and (\ref{eq-MC}) can be equivalently written as
\[\displaystyle \widehat{\bM}_R = \frac{1}{pqT}\sum_{t=1}^{T}{\bX_t^{\star} \bX_t^{\star\prime}} \ \  \text{and} \ \  \widehat{\bM}_C = \frac{1}{pqT}\sum_{t=1}^{T}{\bX_t^{\star\prime} \bX_t^{\star}},\]
which can be viewed as the statistics defined on the transformed data $\bX_t^{\star}$.

\subsection{Estimation of Loading Vectors}\label{sec2.2}
\justifying
\setstretch{1.5}

After obtaining the estimated factor matrix $\wh{\bF}_t$, we proceed to estimate the loading vectors $\wh{\ba}$ and $\wh{\bb}$ in model (\ref{eq-model}). Note that
\[
\begin{aligned}
y_{t+h} &= \ba^{\prime}\bF_t\bb + e_{t+h} \\
        &= \ba^{\prime}\bH_R \bH_R^{-1} \bF_t \bH_C^{-1\prime} \bH_C^{\prime} \bb + e_{t+h} \\
        &= (\bH_R^{\prime} \ba)^{\prime} \bH_R^{-1} \bF_t \bH_C^{-1\prime} (\bH_C^{\prime} \bb) + e_{t+h},
\end{aligned}
\]
which implies that $\wh{\ba}$ and $\wh{\bb}$ are consistent with $\bH_R^{\prime} \ba$ and $\bH_C^{\prime} \bb$, respectively, given that $\wh{\bF}_t$ is consistent with $\bH_R^{-1} \bF_t \bH_C^{-1\prime}$. Hence, we need establish the consistency of the estimators $\wh{\ba}$ and $\wh{\bb}$ by showing that $\Vert \wh{\ba} - \bH_R^{\prime} \ba \Vert_2$ and $\Vert \wh{\bb} - \bH_C^{\prime} \bb \Vert_2$ converge to zero as $p$, $q$ and $T$ go to infinity.

In this paper, we adopt the \textit{LSE} method proposed by \cite{chen2021autoregressive} for estimation. Specifically, the procedure involves solving the following least squares problem:
\begin{equation}
\underset{\wh\ba, \ \wh\bb}{\min}\sum_{t}{\left(y_{t+h} - \wh\ba' \wh{\bF}_t \wh\bb \right)^2}.
\label{eq-arg}
\end{equation}

By applying the first-order condition on (\ref{eq-arg}), we obtain that
\begin{equation}
\displaystyle\sum_{t}y_{t+h}\wh{\bF}_t\wh\bb - \sum_{t}\wh{\bF}_t\wh\bb\wh\bb^\prime \wh{\bF}_t^\prime\wh\ba = \boldsymbol{0},
\label{eq-foc-1}
\end{equation}
\begin{equation}
\displaystyle\sum_{t}y_{t+h}\wh{\bF}_t^\prime\wh\ba - \sum_{t}\wh{\bF}_t^\prime\wh\ba\wh\ba^\prime \wh{\bF}_t\wh\bb = \boldsymbol{0}.
\label{eq-foc-2}
\end{equation}

To solve Equations (\ref{eq-foc-1}) and (\ref{eq-foc-2}), we apply an iterative procedure to update the two loading vectors $\wh\ba$ and $\wh\bb$. Specifically, we alternately update one vector while holding the other fixed, starting from some initial values of $\wh\ba$ and $\wh\bb$. According to Equation (\ref{eq-foc-1}), the update step for $\wh\ba$ given $\wh\bb$ is as follows:
\[\wh\ba \leftarrow \left(\sum_{t}\wh{\bF}_t \wh\bb \wh\bb^\prime \wh{\bF}_t^\prime\right)^{-1} \left(\sum_{t}y_{t+h}\wh{\bF}_t\wh\bb\right).\]

Similarly, according to Equation (\ref{eq-foc-2}), the update step for $\wh\bb$ given $\wh\ba$ is as follows:
\[\wh\bb \leftarrow \left(\sum_{t}\wh{\bF}_t^\prime \wh\ba \wh\ba^\prime \wh{\bF}_t\right)^{-1} \left(\sum_{t}y_{t+h} \wh{\bF}_t^\prime\wh\ba\right).\]

We refer to the estimators $\wh{\ba}$ and $\wh{\bb}$ as the iterative least-squares estimators.

\subsection{Estimation of Latent Dimensions with Factor Matrix}\label{subsec2.3}
In the estimation procedure described above, the factor dimensions \( k \) and \( r \) are pre-specified. However, in practical applications involving empirical data, these dimensions must first be estimated. Following the approach outlined by \cite{chen2023statistical}, our estimation procedure of factor dimensions is conducted using the eigenvalue ratio-based estimator proposed by \cite{lam2012factor}.

Let \( \widehat{\lambda}_1 \geq \widehat{\lambda}_2 \geq \cdots \geq \widehat{\lambda}_k \geq 0 \) represent the ordered eigenvalues of the matrix \( \widehat{\bM}_R \). The ratio-based estimator for determining \( k \) is defined as
\begin{equation}\label{eqn7}
    \widehat{k} = \underset{1 \leq j \leq k_{\max}}{\operatorname*{\arg\max}} \frac{\widehat{\lambda}_j}{\widehat{\lambda}_{j+1}},
\end{equation}
where \( k_{\max} \) is a pre-specified upper bound. In practice, we may choose \( k_{\max} = \lceil p/2 \rceil \), where $\lceil \cdot \rceil$ denotes the ceiling function, as suggested by \cite{lam2012factor}. If the ratio estimator yields only one local maximum, we substitute the second-largest ratio as a stability measure in Equation (\ref{eqn7}). Similarly, the estimator for \( r \), denoted as \( \widehat{r} \), is defined analogously using the eigenvalues of \( \widehat{\bM}_C \).

In this research, we apply the aforementioned method to estimate the dimensions of the factor matrix \( \widehat{\boldsymbol{F}}_t \). For a more comprehensive discussion of the estimation procedure, see \cite{lam2012factor} and \cite{ahn2013eigenvalue} for reference. 

\section{Theoretical Properties}\label{sec3}
\justifying
\setstretch{1.5}

In this section, we establish the theoretical properties of our estimation as $p,q,T\to\infty$. Fisrt, we start with the assumptions that are essential for the theoretical development of our model.

\begin{assumption}\label{A1}
$\alpha$-mixing. Assume $\left(\mathrm{vec}(\bF_t), \mathrm{vec}(\bE_t), \varepsilon_t\right)$ are $\alpha$-mixing. Specifically, a random variable series $\displaystyle\{x_t: \ t=0, \ \pm 1, \ \pm 2, \ \cdots\}$ or a random vector series $\displaystyle\{\bx_t: \ t=0, \ \pm 1, \ \pm 2, \ \cdots\}$ is $\alpha$-mixing if
\[\displaystyle\sum_{h=1}^\infty\alpha(h)^{1-2/\gamma}<\infty,\]
for some $\gamma>2$, where $\displaystyle\alpha(h)=\underset{j\in\mathbb{Z}}{\sup}\underset{A\in\mF_{-\infty}^j,B\in\mF_{j+h}^\infty}{\sup}\vert P(A \cap B)-P(A)P(B)\vert$, and $\mF_{j}^{\ell}$ is the $\sigma$-field generated by $\displaystyle\{x_t: j \leq t \leq \ell\}$ and $\displaystyle\{\bx_t: j \leq t \leq \ell\}$, respectively.
\end{assumption}

\begin{assumption}\label{A2}
Factor and noise matrices. There exist a positive constant $C<\infty$ such that for all $T$, \par
\vspace{0.5em}
1. Factor matrix $\bF_t$ is of fixed dimension $k\times r$ and $\displaystyle\mE{\Vert \bF_t \Vert}^4 \leq C$. \par
2. For all $i\in\left[p\right]$, $j\in\left[q\right]$ and $t\in\left[T\right]$, $\mE\left[e_{t,ij}\right]=0$ and $\mE{\vert e_{t,ij}\vert}^8\leq C$ for $e_{t,ij}$ as entries of $\bE_t$, and $\mE\left[\varepsilon_t\right]=0$, $\mE{\vert \varepsilon_t\vert}^8\leq C$ and $\mathbb{V}\left[\varepsilon_t\right]=\sigma^2$ for $\varepsilon_t$ in model (3). \par
3. Factor and noise are uncorrelated, that is, $\mE{[e_{t,ij}f_{s,lh}]}=0$ for any $t,s\in[T]$, $i\in[p]$, $j\in[q]$, $l\in[k]$, and $h\in[r]$.
\end{assumption}

\begin{assumption}\label{A3}
Loading matrices and vectors. For each row of $\bR$, $\Vert \bR_{i\cdot} \Vert = \mO(1)$, and as $p,q \rightarrow \infty$, we have $\displaystyle \Vert p^{-1} \bR^{\prime} \bR - \boldsymbol\Omega_R \Vert \rightarrow 0$ for some $k \times k$ positive definite matrix $\boldsymbol\Omega_R$. For each row of $\bC$, $\Vert \bC_{i\cdot} \Vert = \mO(1)$, and as $p,q \rightarrow \infty$, we have $\displaystyle \Vert q^{-1} \bC^{\prime} \bC - \boldsymbol\Omega_C \Vert \rightarrow 0$ for some $r \times r$ positive definite matrix $\boldsymbol\Omega_C$. For $\ba$ and $\bb$, $\Vert \ba \Vert = \mO(1)$, $\Vert \bb \Vert = \mO(1)$. Moreover, we impose some constraints on the values of our loading matrices and estimators, i.e., \(\displaystyle\frac{1}{p} \bR^\prime \bR = \bI_k\), \(\displaystyle\frac{1}{q} \bC^\prime \bC = \bI_r\), \(\Vert \wh{\ba} \Vert_2 = 1\), and \(\Vert \ba^\prime \bH_R \Vert_2 = 1\).
\end{assumption}

\begin{assumption}\label{A4}
Cross row (column) correlation of noise $\bE_t$. There exists some positive constant $C<\infty$ such that \par
\vspace{0.5em}
1. Let $\displaystyle\mathbf{U}_E=\mE\left[\frac{1}{qT}\sum_{t=1}^T\bE_t\bE_t^{\prime}\right]$, $\displaystyle\mathbf{V}_{E}=\mE\left[\frac{1}{pT}\sum_{t=1}^{T}\bE_{t}^{\prime}\bE_{t}\right]$, we assume that $\Vert\mathbf{U}_E\Vert_1\leq C$ and $\Vert\mathbf{V}_E\Vert_1\leq C$.\par
2. For all row $i\in[p]$, column $j\in[q]$ and $t\in[T]$, we assume \[\displaystyle\sum_{l\in[p], l \neq i}\sum_{h\in[q], h\neq j}\left|\mE\left[e_{t,ij}e_{t,lh}\right]\right|\leq C.\] \par
3. For any row $i,l\in[p]$, any time $t\in[T]$, and any column \(j\in[q]\),
\[\displaystyle \sum_{m\in [p]} \sum_{s\in [T]} \sum_{h\in [q], h\neq j} \vert \mathrm{cov}[e_{t,ij}e_{t,lj}, \ e_{s,ih}e_{s,mh} ] \vert \leq C.\]

Similarly, for any column $j,h \in [q]$, any time $t \in [T]$, and any row $i \in [p]$,
\[\displaystyle \sum_{m\in [q]} \sum_{s\in [T]} \sum_{l\in [p], l\neq i} \vert \mathrm{cov}[e_{t,ij}e_{t,ih}, \ e_{s,lj}e_{s,lm} ] \vert \leq C.\]
\end{assumption}

\begin{assumption}\label{A5}
(Other model assumptions.) There exists $m>2$, $1<a,b<\infty$, $1/a + 1/b = 1$, such that, for some positive $C<\infty$,

\vspace{0.5em}
1. For any $l\in[k], i\in[p]$ and $t\in[T]$, $\mathbb{E}[{\vert 1/\sqrt{q}\sum_{j=1}^q{e_{t, ij}} \vert}^{mb}] = \mO(1)$, $\mathbb{E}[{\Vert 1/\sqrt{q}\sum_{j=1}^q{\bC_{j\cdot} e_{t, ij}} \Vert}^{mb}] = \mO(1)$, and $\mathbb{E}[{\Vert \mathbf{f}_{t, l\cdot} \Vert}^{ma}]\leq C$. \par
2. For any $h\in[r], j\in[q]$ and $t\in[T]$, $\mathbb{E}[{\vert 1/\sqrt{p}\sum_{i=1}^p{e_{t, ij}} \vert}^{mb}] = \mO(1)$, $\mathbb{E}[{\Vert 1/\sqrt{p}\sum_{i=1}^p{\bR_{i\cdot} e_{t, ij}} \Vert}^{mb}] = \mO(1)$, and $\mathbb{E}[{\Vert \mathbf{f}_{t, \cdot h} \Vert}^{ma}]\leq C$. \par
3. For any $t\in[T]$, $\mathbb{E}[{\vert 1/\sqrt{pq}\sum_{i=1}^p\sum_{j=1}^q{e_{t, ij}} \vert}^{mb}] = \mO(1)$ and $\mathbb{E}[{\Vert 1/\sqrt{pq}\sum_{i=1}^p\sum_{j=1}^q{\bR_{i\cdot} \bC_{j\cdot}^{\prime} e_{t, ij}} \Vert}^{mb}] = \mO(1)$.
\end{assumption}

In addition, to obtain the convergence rate of $\wh\bF_t$, we introduce several covariance matrices involving $\bF_t$, $\bR$ and $\bC$. Let $\boldsymbol{\mu}_F = \mathbb{E}[\bF_t]$ and
\[\displaystyle\boldsymbol\Sigma_{FC} := \mathbb{E}\left[ (\bF_t - \boldsymbol{\mu}_F) (\bC^{\prime} \bC/q) (\bF_t - \boldsymbol{\mu}_F)^{\prime} \right],\]
\[\displaystyle\boldsymbol\Sigma_{FR} := \mathbb{E}\left[ (\bF_t - \boldsymbol{\mu}_F)^{\prime} (\bR^{\prime} \bR/p) (\bF_t - \boldsymbol{\mu}_F) \right],\]
then
\[\displaystyle\widetilde{\boldsymbol\Sigma}_{FC} := \frac{1}{q}\mathbb{E}\left[ \widetilde{\bF}_t \bC^{\prime} \bC \widetilde{\bF}_t^{\prime} \right] = \boldsymbol\Sigma_{FC} + (\alpha + 1) \frac{1}{q} \boldsymbol{\mu}_F \bC^{\prime} \bC \boldsymbol{\mu}_F^{\prime},\]
\[\displaystyle\widetilde{\boldsymbol\Sigma}_{FR} := \frac{1}{p}\mathbb{E}\left[ \widetilde{\bF}_t^{\prime} \bR^{\prime} \bR \widetilde{\bF}_t \right] = \boldsymbol\Sigma_{FR} + (\alpha + 1) \frac{1}{p} \boldsymbol{\mu}_F^{\prime} \bR^{\prime} \bR \boldsymbol{\mu}_F.\]
\indent The matrix $\boldsymbol\Sigma_{FC}$ can be interpreted as the row covariance of $\bF_t$ scaled by the strength of the column variances in $\bF_t\bC^{\prime}$, while $\boldsymbol\Sigma_{FR}$ represents the column covariance of $\bR\bF_t^{\prime}$ scaled by the strength of the row variances in $\bR\bF_t^{\prime}$. These matrices, $\boldsymbol\Sigma_{FC}$ and $\boldsymbol\Sigma_{FR}$, capture the aggregated moment information of the rows and columns in $\bF_t\bC^{\prime}$ and $\bR\bF_t^{\prime}$, respectively.

\begin{assumption}\label{A6}
(Distinct eigenvalues conditions.) The eigenvalues of the $k \times k$ matrix $\boldsymbol\Omega_R\widetilde{\boldsymbol\Sigma}_{FC}$ are distinct and so are the eigenvalues of the $r \times r$ matrix $\boldsymbol\Omega_C\widetilde{\boldsymbol\Sigma}_{FR}$.
\end{assumption}

Now, we are ready to present the theoretical properties of our estimators. To facilitate the analysis, we first introduce the auxiliary matrices $\bH_{R}$, $\bH_{C}$, $\bV_{R, pqT}$ and $\bV_{C, pqT}$, as defined in \cite{chen2023statistical}. As previously mentioned, the matrices $\bR$, $\bC$, and $\bF_t$ are not separately identifiable. We demonstrate in the following that, for any ground truth $\bR$, $\bC$, and $\bF_t$, there exist invertible matrices $\bH_R$ and $\bH_C$ such that our estimators our estimators $\wh\bR$ and $\wh\bC$ are consistent for $\bR\bH_R$ and $\bC\bH_C$, respectively, and ${\wh\bF}_t$ is a consistent estimator of the rotated factor matrix $\bH_R^{-1}\bF_t\bH_C^{-1\prime}$.

Let $\bV_{R,pqT}\in\mathbb{R}^{k\times k}$ and $\bV_{C,pqT}\in\mathbb{R}^{r\times r}$ be diagonal matrices consisting of the top $k$ and $r$ largest eigenvalues which are in descending order of $\displaystyle\wh{\bM}_R = \frac{1}{pqT}\sum_{t=1}^{T}{\bX^{\star}_t \bX^{\star\prime}_t}$ and $\displaystyle\wh{\bM}_C = \frac{1}{pqT}\sum_{t=1}^{T}{{\bX^{\star\prime}_t \bX^{\star}_t}}$, respectively. According to the definition of the estimators $\wh\bR$ and $\wh\bC$, we have
\[\displaystyle\wh\bR = \frac{1}{pqT}\sum_{t=1}^{T}{\bX^{\star}_t \bX^{\star\prime}_t}\wh\bR\bV_{R,pqT}^{-1}, \ \wh\bC = \frac{1}{pqT}\sum_{t=1}^{T}{\bX^{\star\prime}_t \bX^{\star}_t}\wh\bC\bV_{C,pqT}^{-1}.\]

Define $\bH_R\in\mathbb{R}^{k\times k}$ and $\bH_C\in\mathbb{R}^{r\times r}$ as 
\[\bH_R = \frac{1}{pqT}\sum_{t=1}^{T}{\widetilde{\bF}_t \bC^{\prime} \bC \widetilde{\bF}_t^{\prime} \bR^{\prime} \wh\bR \bV_{R, pqT}^{-1}} \in \mathbb{R}^{k\times k},\]
\[\bH_C = \frac{1}{pqT}\sum_{t=1}^{T}{\widetilde{\bF}_t^{\prime} \bR^{\prime} \bR \widetilde{\bF}_t \bC^{\prime} \wh\bC \bV_{C, pqT}^{-1}} \in \mathbb{R}^{r\times r},\]
which are bounded as $p,q,T \rightarrow \infty$.

The following propositions, which have been proven in \cite{chen2023statistical}, provide the convergence rates for \(\wh{\bR}\), \(\wh{\bC}\), and \(\wh{\bF}\).

\vspace{0.5em}
\begin{proposition}\label{P1}
Under Assumptions \ref{A1}-\ref{A5}, we have, as $k, r$ fixed and $p, q, T \rightarrow \infty$,
\[\displaystyle \frac{1}{p}{\Vert \wh\bR - \bR\bH_R \Vert}_F^{2} = \mathcal{O}_p\left(\frac{1}{\min\{p, qT\}}\right),\]
\[\displaystyle \frac{1}{q}{\Vert \wh\bC - \bC\bH_C \Vert}_F^{2} = \mathcal{O}_p\left(\frac{1}{\min\{q, pT\}}\right).\]

Consequently,
\[\displaystyle \frac{1}{p}{\Vert \wh\bR - \bR\bH_R \Vert}^{2} = \mathcal{O}_p\left(\frac{1}{\min\{p, qT\}}\right),\]
\[\displaystyle \frac{1}{q}{\Vert \wh\bC - \bC\bH_C \Vert}^{2} = \mathcal{O}_p\left(\frac{1}{\min\{q, pT\}}\right).\]
\end{proposition}

\vspace{0.5em}
\begin{proposition}\label{P2}
Under Assumptions \ref{A1}-\ref{A6}, we have, as $k,r$ fixed and $p,q,T \rightarrow \infty$, we have 
\[{\Vert {\wh\bF}_t - \bH_R^{-1}\bF_t\bH_C^{-1\prime} \Vert}_F^2 = \mathcal{O}_p\left( \frac{1}{\min{\{ p, q \}}} \right).\]
\end{proposition}

\vspace{0.5em}
\begin{remark}
This part of the proof follows from the theoretical results established in \cite{chen2023statistical}, and is not elaborated further here.
\end{remark}

\vspace{0.5em}
Now, based on all the conditions and propositions above, we can finally derive the convergence rates as well as the asymptotic normality properties for our estimators \(\wh{\ba}\) and \(\wh{\bb}\).

\vspace{0.5em}
\begin{theorem}\label{T1}
Under Assumptions \ref{A1}-\ref{A6}, as \(k, r\) fixed and \(p, q, T \to \infty\), we have

\[{\Vert \wh\ba - \bH_R^{\prime} \ba \Vert}_2 = \mathcal{O}_p\left(\frac{1}{\sqrt{T}}\right),\]
\[{\Vert \wh\bb - \bH_C^{\prime} \bb \Vert}_2 = \mathcal{O}_p\left(\frac{1}{\sqrt{T}}\right).\]

\vspace{0.5em}
Furthermore, define $\boldsymbol{\gamma} := (\ba^{\prime}\bH_R, \boldsymbol{0}^{\prime})^{\prime}$, $\boldsymbol{U}_t = \begin{pmatrix} \displaystyle\bH_R^{-1} \bF_t \bb \\ \displaystyle\bH_C^{-1} \bF_t^{\prime} \ba \end{pmatrix}$, $\boldsymbol{H} := \mathbb{E}[\boldsymbol{U}_t \boldsymbol{U}_t^{\prime}] + \boldsymbol{\gamma}\boldsymbol{\gamma}^{\prime}$, we have the asymptotic normality properties with $\begin{pmatrix} \wh\ba \\ \wh\bb \end{pmatrix}$ and $\wh\bb \otimes \wh\ba$, which is
\[\displaystyle \sqrt{T}\begin{pmatrix} \widehat{\ba} - \bH_R^{\prime}\ba \\  \widehat{\bb} - \bH_C^{\prime}\bb \end{pmatrix} \Rightarrow \mathcal{N}\left(\boldsymbol{0}, \boldsymbol{\Xi} \right),\]
\[\displaystyle \sqrt{T} \left[ \wh\bb \otimes \wh\ba - \left(\bH_C^{\prime}\bb\right) \otimes \left(\bH_R^{\prime}\ba\right) \right] \Rightarrow \mathcal{N}\left(\boldsymbol{0}, \boldsymbol{V} \boldsymbol{\Xi} \boldsymbol{V}^{\prime} \right),\]
where $\displaystyle \boldsymbol{\Xi} := \sigma^2 \bH^{-1} \mathbb{E}[\boldsymbol{U}_t \boldsymbol{U}_t^{\prime}] \bH^{-1}$, $\displaystyle \boldsymbol{V} := \left[\bb^{\prime}\bH_C \otimes \boldsymbol{I},\boldsymbol{I} \otimes \bH_R^{\prime}\ba \right]$.
\end{theorem}

\vspace{0.5em}
\begin{remark}\label{R3}
This result is similar to Theorem 3 in \cite{chen2021autoregressive}, and it characterizes the asymptotic properties of the estimators $\wh\ba$ and $\wh\bb$ in both the vectorized form and the Kronecker product form, with both estimators achieving the standard $\sqrt{T}$-consistency. Importantly, since the loading vectors are estimated using the Least Squares Estimation (LSE) method after the factor estimation step via $\alpha$-PCA, the input data for the LSE procedure does not consist of the true factor process $\bF_t$, but the rotation version $\wh\bF_t$.
\end{remark}

\section{A Supervised Screening Refinement}\label{sec4}
\justifying
\setstretch{1.5}

Through the application of factor dimensionality reduction, we have derived the latent factor matrix ${\wh\bF}_t$. However, the current procedure only accounts for the structural information contained in the predictor $\bX_t$, without leveraging the potentially valuable information embedded in the target variable $y_t$. In the process of handling high-dimensional data, numerous studies have employed screening methods, such as variable selection techniques, to enhance model estimation or prediction accuracy, as demonstrated in the works of \cite{li2012feature}, \cite{shao2014martingale}, and \cite{guo2023threshold}, among others. In light of the aforementioned approach, given the correlation between $y_t$ and the predictor $\bX_t$,  which may offer informative signals for forecasting, we proposed a new supervised data screening procedure. This ensures that the dataset retains only the most relevant predictors for the forecasting task. Specifically, we suggest that $\bX_t$ should be trained by $y_t$ before the factor extraction procedure is applied.

In consideration of the original model structures in (\ref{eq-model}), our objective is to integrate the correlation between the observation matrix $\bX_t$ and the target variable $y_t$ into the model. To quantitatively evaluate this relationship, we compute a correlation matrix $\bP$ where each entry $\rho_{ij}$ represents the correlation coefficient between $y_t$ and the $(i,j)$-th element of the observation matrix $\bX_t$, denoted as $x_{ij,t}$. We then calculate the average correlation for each row and column, denoted respectively as $\bar\rho_i$ and $\bar\rho_j$. Using a pre-specified threshold parameter $\rho_\delta$, we filter out the $i$-th row and $j$-th column of $\bX_t$ if $\bar\rho_i<\rho_\delta$ or $\bar\rho_j<\rho_\delta$. This filtering procedure removes rows and columns that exhibit weak associations with the target variable, thereby reducing noise and potentially enhancing the forecasting performance. The screening algorithm can be formally expressed as follows.

\begin{algorithm}[!ht]
    \setstretch{1.25}
    \renewcommand{\algorithmicrequire}{\textbf{\textrm{Input:}}} 
    \renewcommand{\algorithmicensure}{\textbf{\textrm{Output:}}}
    \caption{\textbf{\textrm{Calculation of the Refined Observation Matrix \(\widetilde{\bX}_t\)}}}
    \begin{algorithmic}
        \vspace*{10pt}
        \REQUIRE Original observation matrix \(\bX_t\), target time series \(y_t\).
        \vspace*{10pt}
        \ENSURE Refined observation matrix \(\widetilde{\bX}_t\).
        \vspace*{10pt}
        \STATE \textbf{Step 1.} Compute the correlation coefficients between each element \(x_{ij,t}\) of the observation matrix \(\bX_t\) and the target variable \(y_t\), and construct the correlation matrix \(\bP\), where each entry is defined as \(\rho_{ij} = \mathrm{corr}(x_{ij,t}, y_t)\);
        \vspace*{10pt}
        \STATE \textbf{Step 2.} Calculate the average correlation coefficients for each row and column of \(\bP\), denoted as
        \[
        \bar{\rho}_i = \frac{1}{n} \sum_{j=1}^{n} |\rho_{ij}| \quad \text{and} \quad \bar{\rho}_j = \frac{1}{m} \sum_{i=1}^{m} |\rho_{ij}|,
        \]
        respectively, where \(m\) and \(n\) represent the number of rows and columns of the matrix \(\bP\);
        \vspace*{10pt}
        \STATE \textbf{Step 3.} Remove the \(i\)-th row and \(j\)-th column of the original matrix \(\bX_t\) if the corresponding average correlation coefficients \(\bar{\rho}_i\) or \(\bar{\rho}_j\) are smaller than a predetermined threshold parameter \(\rho_\delta\), and obtain the refined matrix \(\widetilde{\bX}_t\). Formally, define the set of retained indices as
        \[
        \widehat{\Delta} := \left\{ (i, j) \mid 1 \leq i \leq m, \ 1 \leq j \leq n: \bar{\rho}_i \geq \rho_\delta \ \text{and} \ \bar{\rho}_j \geq \rho_\delta \right\},
        \]
        such that the refined observation matrix \(\widetilde{\bX}_t\) consists of the elements \(x_{\widehat{\Delta}, t}\).
        \vspace*{10pt}
    \end{algorithmic}
\end{algorithm}

After this screening procedure, we obtain a refined observation matrix \(\widetilde{\bX}_t \in \mathbb{R}^{\widetilde{p} \times \widetilde{q}}\) with a reduced dimension \(\widetilde{p} \times \widetilde{q}\), which has been trained based on the information from \(y_t\). We then proceed with \textit{$\alpha$-PCA} and the \textit{LSE} method as previously outlined. Empirically, the screening refinement process leads to a reduction in forecasting loss compared to using the original observation matrix \(\bX_t\), which mainly stems from the optimized matrix obtained through the selection process. This refined matrix better captures the relationship between the target variable \(y_t\) and \(\bX_t\), thereby enhancing the model's performance by reducing noise, improving signal quality, and decreasing model complexity. These results indicate that our proposed screening method can effectively reduce forecasting error and enhance prediction accuracy.

\section{Simulation}\label{sec5}
\justifying
\setstretch{1.5}

In this section, we conduct Monte Carlo simulations to assess the finite-sample performance of the estimators $\wh\ba$, $\wh\bb$, and $\wh\bF_t$, and evaluate the consistency of $\wh\bF_t$ and $\wh\bb \otimes \wh\ba$, as well as the asymptotic normality of $\begin{pmatrix} \wh\ba \\ \wh\bb \end{pmatrix}$ and $\wh\bb \otimes \wh\ba$ under various data generating processes (DGPs). The following subsection outlines the model settings used in the simulation.

\subsection{Settings}\label{sec5.1}
\justifying
\setstretch{1.5}

The specific steps of the simulation in this article are as follows. \par

\vspace{0.5em}
1. \textit{Generation of the factor matrix $\boldsymbol{F}_t$.} Let $\boldsymbol{F}_t \in \mathbb{R}^{k_1 \times k_2}$, where $k_1 = 3$ and $k_2 = 2$. The matrix $\boldsymbol{F}_t$ is generated using two methods. (i) \textit{Matrix Normal Distribution.} Let $\boldsymbol{F}_t \sim \mathcal{M}\mathcal{N}(\boldsymbol{\mu_F}, \boldsymbol{U}, \boldsymbol{V})$, where $\boldsymbol{\mu_F} \in \mathbb{R}^{k_1 \times k_2}$ is a matrix with all zero entries, $\boldsymbol{U} \in \mathbb{R}^{k_1 \times k_1}$, and $\boldsymbol{V} \in \mathbb{R}^{k_2 \times k_2}$ are identity matrices. (ii) \textit{First-order Matrix Autoregression (MAR(1)).} The factor matrix $\boldsymbol{F}_t$ follows the first-order matrix autoregression time-series process 
\(
\boldsymbol{F}_t = \boldsymbol{\varPhi}_1 \boldsymbol{F}_{t-1} \boldsymbol{\varPhi}_2^{\prime} + \boldsymbol{E}_t
\)
for $t = 1, \dots, T$, where the autoregression coefficient matrices $\boldsymbol{\varPhi}_1 \in \mathbb{R}^{k_1 \times k_1}$ and $\boldsymbol{\varPhi}_2 \in \mathbb{R}^{k_2 \times k_2}$ are diagonal matrices with diagonal elements sampled from the uniform distribution $\mathcal{U}(-1, 1)$. This setup ensures the stationarity and causality of the process $\boldsymbol{F}_t$ (see Proposition 1, \cite{chen2021autoregressive}). The elements of the noise matrix $\boldsymbol{E}_t$ are drawn from $\mathcal{N}(0, 1)$, and the initial values of the entries of $\boldsymbol{F}_0$ are set to zero.

\vspace{0.5em}
2. \textit{Generation of the observation matrix $\boldsymbol{X}_t$.} Given the generated factor matrix $\boldsymbol{F}_t$, the observation matrix $\boldsymbol{X}_t \in \mathbb{R}^{p \times q}$ is generated as follows, which is \(\boldsymbol{X}_t = \boldsymbol{R} \boldsymbol{F}_t \boldsymbol{C}^{\prime} + \boldsymbol{E}_t,\)
where the entries of $\boldsymbol{R} \in \mathbb{R}^{p \times k_1}$ and $\boldsymbol{C} \in \mathbb{R}^{q \times k_2}$ are independently drawn from the uniform distribution $\mathcal{U}(-1, 1)$. The noise matrix $\boldsymbol{E}_t \in \mathbb{R}^{p \times q}$ is generated using one of the following three methods, as described in \cite{chen2023statistical}. (i) \textit{Uncorrelated.} Assume $\boldsymbol{E}_t \sim \mathcal{MN}(\boldsymbol{0}, \boldsymbol{I}, \boldsymbol{I})$. In this case, the elements of $\boldsymbol{E}_t$ are uncorrelated across rows, columns, and time. (ii) \textit{Weakly correlated across time.} The noise matrix $\boldsymbol{E}_t$ follows the MAR(1) process
\(
\boldsymbol{E}_t = \boldsymbol{\varPsi}_1 \boldsymbol{E}_{t-1} \boldsymbol{\varPsi}_2^{\prime} + \boldsymbol{U}_t,
\)
where the autoregression coefficient matrices $\boldsymbol{\varPsi}_1 \in \mathbb{R}^{p \times p}$ and $\boldsymbol{\varPsi}_2 \in \mathbb{R}^{q \times q}$ are specified in accordance with the settings of $\boldsymbol{\varPhi}_1$ and $\boldsymbol{\varPhi}_2$ described earlier. The entries of the noise matrix $\boldsymbol{U}_t$ are drawn from $\mathcal{N}(0, 1)$ independently across time. This configuration results in elements of $\boldsymbol{E}_t$ that are independent at each time point, but exhibit weak correlation across time. (iii) \textit{Weakly correlated across rows and columns.} Assume that $\boldsymbol{E}_t \sim \mathcal{MN}(\boldsymbol{0}, \boldsymbol{U_E}, \boldsymbol{V_E})$, where the diagonal elements of $\boldsymbol{U_E}$ and $\boldsymbol{V_E}$ are set to $1$, and the off-diagonal elements are set to $1/p$ and $1/q$, respectively. Under this setting, $\boldsymbol{E}_t$ is uncorrelated across time. However, due to the non-diagonal structure of $\boldsymbol{U_E}$ and $\boldsymbol{V_E}$, the entries of $\boldsymbol{E}_t$ exhibit weak correlation across rows and columns.

\vspace{0.5em}
3. \textit{Generation of the target response variables $y_{t+h}$.} Simultaneously with the generation of $\boldsymbol{F}_t$, the target response variables $y_{t+h}$ is generated by
\(
y_{t+h} = \ba^{\prime} \boldsymbol{F}_t \bb + e_{t+h},
\)
where $h$ denotes the forecasting step, which is set to $h = 1$. The loading vectors $\ba \in \mathbb{R}^{k_1 \times 1}$ and $\bb \in \mathbb{R}^{k_2 \times 1}$ are constrained such that $\Vert \ba \Vert_2 = 1$ and both elements are drawn independently from the standard normal distribution $\mathcal{N}(0,1)$. The independent and identical white noise term $e_t \sim \mathcal{N}(0,1)$ represents the innovation at the forecasting horizon.

\subsection{Comparison of Convergence}
\justifying
\setstretch{1.5}

To evaluate the performance of different configurations, we consider several settings in our simulation. The dimensions of the observation matrix \(\bX_t\) are set as \((p,q) = (5,10), (10,10), (20,20)\), while the dimensions of the factor matrix \(\bF_t\) are set as \((k, r) = (2,3)\). The time horizon \(T\) is chosen from the set \(T = \{0.5pq, pq, 2pq\}\) in Table \ref{Table.1}, similar to the setup in \cite{wang2019factor}, and \(T = \{100, 200, 400, 5000\}\) in Table \ref{Table.2}, similar to the setup in \cite{chen2021autoregressive}. We use the \(\ell_2\)-norm loss and the logarithmic Frobenius norm loss to compute the estimation errors of the factor matrix \(\wh{\bF}_t\) and the Kronecker product of the loading vectors \(\wh{\bb} \otimes \wh{\ba}\), respectively. These are written as \({\big\Vert \widehat{\boldsymbol{F}}_t - \boldsymbol{H}_R^{-1} \boldsymbol{F}_t \boldsymbol{H}_C^{\prime -1} \big\Vert}_2\) and \(\log \big( \big\Vert \wh{\bb} \otimes \wh{\ba} - \bH_C^{\prime} \bb \otimes \bH_R^{\prime} \ba \big\Vert_F^2 \big)\). We conduct 200 replications for each configuration. The corresponding estimation errors and boxplots under different settings are presented below.

\begin{table}[!ht]
  \centering
  \caption{\setstretch{1.25}Means and standard deviations (in parentheses) of estimation loss of \(\wh\bF_t\). Six distinct experimental settings are considered, derived from all possible combinations of \( \bF_t \sim \text{i}, \text{ii} \) and \( \bE_t \sim \text{i}, \text{ii}, \text{iii} \), as outlined in Section \ref{sec5.1}. Three configurations for the dimensions of the observation matrix \( \bX_t \) are examined: \( (p,q) = (5,10) \), \( (10,10) \), and \( (20,20) \). The time horizon \( T \) is selected from the set \( T = \{0.5pq, pq, 2pq\} \). For simplicity in presenting the estimation results, the dimensions of the latent factor matrix \( \bF_t \) are fixed at \( (k,r) = (3,2) \), and the hyperparameter \(\alpha\) in \textit{\(\alpha\)-PCA} method is set with \(\alpha=0\). The experiments are based on 200 replications to ensure robust statistical estimates.}
  \vspace{0.5em}
  
  \setstretch{1.25}
  \footnotesize
    \begin{tabular}{cccccccccc}
    \toprule
          & \multicolumn{3}{c}{$F_t\sim \text{i}$, $E_t\sim \text{i}$} & \multicolumn{3}{c}{$F_t\sim \text{i}$, $E_t\sim \text{ii}$} & \multicolumn{3}{c}{$F_t\sim \text{i}$, $E_t\sim \text{iii}$} \\
          \((p,q)\) & \multicolumn{1}{c}{\(T=0.5pq\)} & \multicolumn{1}{c}{\(T=pq\)} & \multicolumn{1}{c}{\(T=2pq\)} & \multicolumn{1}{c}{\(T=0.5pq\)} & \multicolumn{1}{c}{\(T=pq\)} & \multicolumn{1}{c}{\(T=2pq\)} & \multicolumn{1}{c}{\(T=0.5pq\)} & \multicolumn{1}{c}{\(T=pq\)} & \multicolumn{1}{c}{\(T=2pq\)} \\
\cmidrule{2-10}    \multirow{2}{*}{(5,10)} & 0.162 & 0.099 & 0.075 & 0.243 & 0.162 & 0.107 & 0.173 & 0.132 & 0.090 \\
          & (0.050) & (0.034) & (0.025) & (0.133) & (0.083) & (0.054) & (0.069) & (0.060) & (0.032) \\
    \multirow{2}{*}{(10,10)} & 0.092 & 0.066 & 0.045 & 0.113 & 0.083 & 0.056 & 0.099 & 0.070 & 0.051 \\
          & (0.026) & (0.019) & (0.013) & (0.031) & (0.025) & (0.021) & (0.033) & (0.018) & (0.014) \\
    \multirow{2}{*}{(20,20)} & 0.048 & 0.034 & 0.025 & 0.054 & 0.037 & 0.027 & 0.049 & 0.035 & 0.025 \\
          & (0.011) & (0.008) & (0.006) & (0.013) & (0.008) & (0.006) & (0.012) & (0.007) & (0.006) \\
\cmidrule{2-10}          & \multicolumn{3}{c}{$F_t\sim \text{ii}$, $E_t\sim \text{i}$} & \multicolumn{3}{c}{$F_t\sim \text{ii}$, $E_t\sim \text{ii}$} & \multicolumn{3}{c}{$F_t\sim \text{ii}$, $E_t\sim \text{iii}$} \\
          \((p,q)\) & \multicolumn{1}{c}{\(T=0.5pq\)} & \multicolumn{1}{c}{\(T=pq\)} & \multicolumn{1}{c}{\(T=2pq\)} & \multicolumn{1}{c}{\(T=0.5pq\)} & \multicolumn{1}{c}{\(T=pq\)} & \multicolumn{1}{c}{\(T=2pq\)} & \multicolumn{1}{c}{\(T=0.5pq\)} & \multicolumn{1}{c}{\(T=pq\)} & \multicolumn{1}{c}{\(T=2pq\)} \\
\cmidrule{2-10}    \multirow{2}{*}{(5,10)} & 0.189 & 0.125 & 0.092 & 0.232 & 0.169 & 0.106 & 0.197 & 0.150 & 0.104 \\
          & (0.069) & (0.046) & (0.037) & (0.104) & (0.090) & (0.046) & (0.082) & (0.062) & (0.042) \\
    \multirow{2}{*}{(10,10)} & 0.108 & 0.079 & 0.059 & 0.127 & 0.091 & 0.065 & 0.131 & 0.080 & 0.060 \\
          & (0.038) & (0.029) & (0.020) & (0.042) & (0.033) & (0.026) & (0.059) & (0.026) & (0.022) \\
    \multirow{2}{*}{(20,20)} & 0.058 & 0.042 & 0.029 & 0.064 & 0.046 & 0.030 & 0.062 & 0.038 & 0.030 \\
          & (0.018) & (0.013) & (0.009) & (0.021) & (0.016) & (0.010) & (0.019) & (0.012) & (0.010) \\
    \midrule
          &       &       &       &       &       &       &       &       &  \\
    \end{tabular}%
  \label{Table.1}%
\end{table}%

Table \ref{Table.1} presents the means and standard deviations of the estimation loss for $\wh\bF_t$. In Table \ref{Table.1}, we report the estimation loss across various configurations, including six combinations of the generation processes for $\bF_t$ and $\bE_t$, three configurations of $(p,q)$, and three settings for the time horizon $T$. From Table \ref{Table.1}, it can be observed that the convergence of the estimation errors improves with smaller variance as $(p,q)$ increases. Additionally, a slight improvement in the convergence of the estimation errors is observed with higher values of $T$, accompanied by a decrease in variance. Furthermore, we find that the estimation errors are smaller for DGP (i) of $\bF_t$ compared to DGP (ii). This is reasonable because the estimation procedure of \(\wh\bF_t\) in this paper relies solely on static contemporaneous information, which is also used in the DGP (i), resulting in better estimation performance. In contrast, DGP (ii) incorporates dynamic information from the previous time step, $t-1$. Other findings are consistent with those reported in \cite{chen2023statistical}, further supporting the consistency property of $\wh\bF_t$ as stated in Proposition \ref{P2}.

Next, we examine the estimation loss for \(\wh\bb\otimes\wh\ba\) using simulated data.

\begin{table}[!ht]
  \centering
  \setstretch{1.25}
  \captionsetup{width=\textwidth}
  \footnotesize
  \caption{\setstretch{1.25}Means and standard deviations (in parentheses) of estimation loss of \(\wh\bb\otimes\wh\ba\). Six distinct experimental settings are considered, derived from all possible combinations of \( \bF_t \sim \text{i}, \text{ii} \) and \( \bE_t \sim \text{i}, \text{ii}, \text{iii} \), as outlined in Section \ref{sec5.1}. Three configurations for the dimensions of the observation matrix \( \bX_t \) are examined: \( (p,q) = (5,10) \), \( (10,10) \), and \( (20,20) \). The time horizon \( T \) is selected from the set \( T = \{100, 200, 400, 5000\} \). For simplicity in presenting the estimation results, the dimensions of the latent factor matrix \( \bF_t \) are fixed at \( (k,r) = (3,2) \), and the hyperparameter \(\alpha\) in \textit{\(\alpha\)-PCA} method is set with \(\alpha=0\). The experiments are based on 200 replications to ensure robust statistical estimates.}
    \vspace{0.5em}
    \begin{tabular}{ccccccccc}
    \toprule
          & \multicolumn{4}{c}{$F_t\sim \text{i}$, $E_t\sim \text{i}$} & \multicolumn{4}{c}{$F_t\sim \text{i}$, $E_t\sim \text{ii}$} \\
          \multicolumn{1}{c}{\((p,q)\)} & \(T=100\) & \(T=200\) & \(T=400\) & \(T=5000\) & \(T=100\) & \(T=200\) & \(T=400\) & \(T=5000\) \\
\cmidrule{2-9}    \multirow{2}{*}{(5,10)} & 0.085 & -0.277 & -0.587 & -1.895 & 0.052 & -0.360 & -0.654 & -1.925 \\
          & (0.441) & (0.371) & (0.416) & (0.489) & (0.418) & (0.434) & (0.460) & (0.502) \\
    \multirow{2}{*}{(10,10)} & 0.113 & -0.263 & -0.682 & -1.883 & 0.128 & -0.227 & -0.553 & -1.831 \\
          & (0.464) & (0.449) & (0.365) & (0.478) & (0.448) & (0.477) & (0.378) & (0.425) \\
    \multirow{2}{*}{(20,20)} & 0.046 & -0.370 & -0.613 & -1.754 & 0.090 & -0.258 & -0.614 & -1.575 \\
          & (0.433) & (0.398) & (0.393) & (0.421) & (0.423) & (0.387) & (0.395) & (0.369) \\
\cmidrule{2-9}          & \multicolumn{4}{c}{$F_t\sim \text{i}$, $E_t\sim \text{iii}$} & \multicolumn{4}{c}{$F_t\sim \text{ii}$, $E_t\sim \text{i}$} \\
          \multicolumn{1}{c}{\((p,q)\)} & \(T=100\) & \(T=200\) & \(T=400\) & \(T=5000\) & \(T=100\) & \(T=200\) & \(T=400\) & \(T=5000\) \\
\cmidrule{2-9}    \multirow{2}{*}{(5,10)} & 0.131 & -0.269 & -0.570 & -1.800 & 0.107 & -0.193 & -0.563 & -1.893 \\
          & (0.400) & (0.375) & (0.388) & (0.414) & (0.456) & (0.449) & (0.427) & (0.411) \\
    \multirow{2}{*}{(10,10)} & 0.173 & -0.207 & -0.498 & -1.889 & -0.037 & -0.290 & -0.621 & -1.803 \\
          & (0.426) & (0.354) & (0.435) & (0.446) & (0.469) & (0.427) & (0.458) & (0.464) \\
    \multirow{2}{*}{(20,20)} & 0.145 & -0.284 & -0.635 & -1.639 & 0.074 & -0.303 & -0.639 & -1.801 \\
          & (0.412) & (0.368) & (0.443) & (0.392) & (0.453) & (0.444) & (0.423) & (0.390) \\
\cmidrule{2-9}          & \multicolumn{4}{c}{$F_t\sim \text{ii}$, $E_t\sim \text{ii}$} & \multicolumn{4}{c}{$F_t\sim \text{ii}$, $E_t\sim \text{iii}$} \\
          \multicolumn{1}{c}{\((p,q)\)} & \(T=100\) & \(T=200\) & \(T=400\) & \(T=5000\) & \(T=100\) & \(T=200\) & \(T=400\) & \(T=5000\) \\
\cmidrule{2-9}    \multirow{2}{*}{(5,10)} & 0.038 & -0.283 & -0.698 & -1.932 & 0.053 & -0.235 & -0.604 & -1.848 \\
          & (0.481) & (0.380) & (0.449) & (0.420) & (0.375) & (0.423) & (0.388) & (0.432) \\
    \multirow{2}{*}{(10,10)} & 0.048 & -0.315 & -0.657 & -1.793 & 0.245 & -0.250 & -0.500 & -1.766 \\
          & (0.410) & (0.390) & (0.523) & (0.444) & (0.483) & (0.411) & (0.423) & (0.446) \\
    \multirow{2}{*}{(20,20)} & -0.053 & -0.283 & -0.708 & -1.732 & 0.025 & -0.378 & -0.650 & -1.785 \\
          & (0.401) & (0.429) & (0.365) & (0.427) & (0.471) & (0.344) & (0.449) & (0.362) \\
    \bottomrule
    \end{tabular}%
  \label{Table.2}%
\end{table}%

Table \ref{Table.2} presents the means and standard deviations of the estimation loss for $\wh\bb \otimes \wh\ba$. In Table \ref{Table.2}, we report the estimation loss under several configurations, similar to the setup in Table \ref{Table.1}, but excluding the time horizon $T$, which is chosen from the set $T = \{100, 200, 400, 5000\}$. This setup is comparable to the one used in \cite{chen2021autoregressive}. From Table \ref{Table.2}, we observe that the estimation errors tend to decrease as $T$ increases, while the errors show little change as $(p,q)$ increases. Furthermore, the estimation results remain stable across various settings, indicating that our estimation procedure is robust. This is consistent with the consistency property of $\wh\bb \otimes \wh\ba$ in Theorem \ref{T1}.

\begin{figure}[!ht]
    \centering
    \begin{minipage}{0.49\linewidth}
    \centering
    \includegraphics[width=0.9\linewidth]{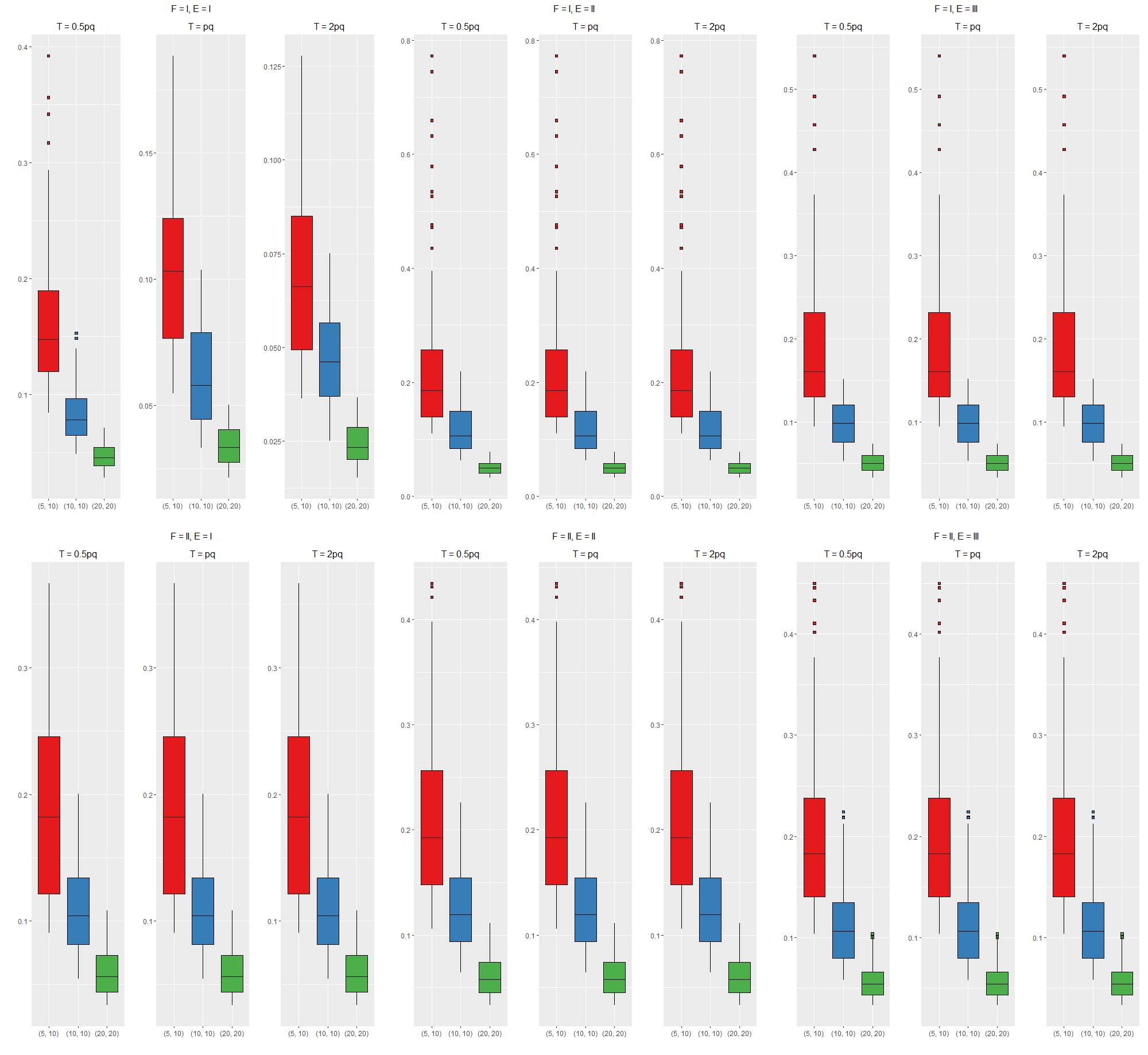}
    \caption{\setstretch{1.25}The boxplot of estimation loss for $\widehat{\boldsymbol{F}}_t$.}
    \label{Figure.1}
    \end{minipage}
    \begin{minipage}{0.49\linewidth}
    \centering
    \includegraphics[width=0.9\linewidth]{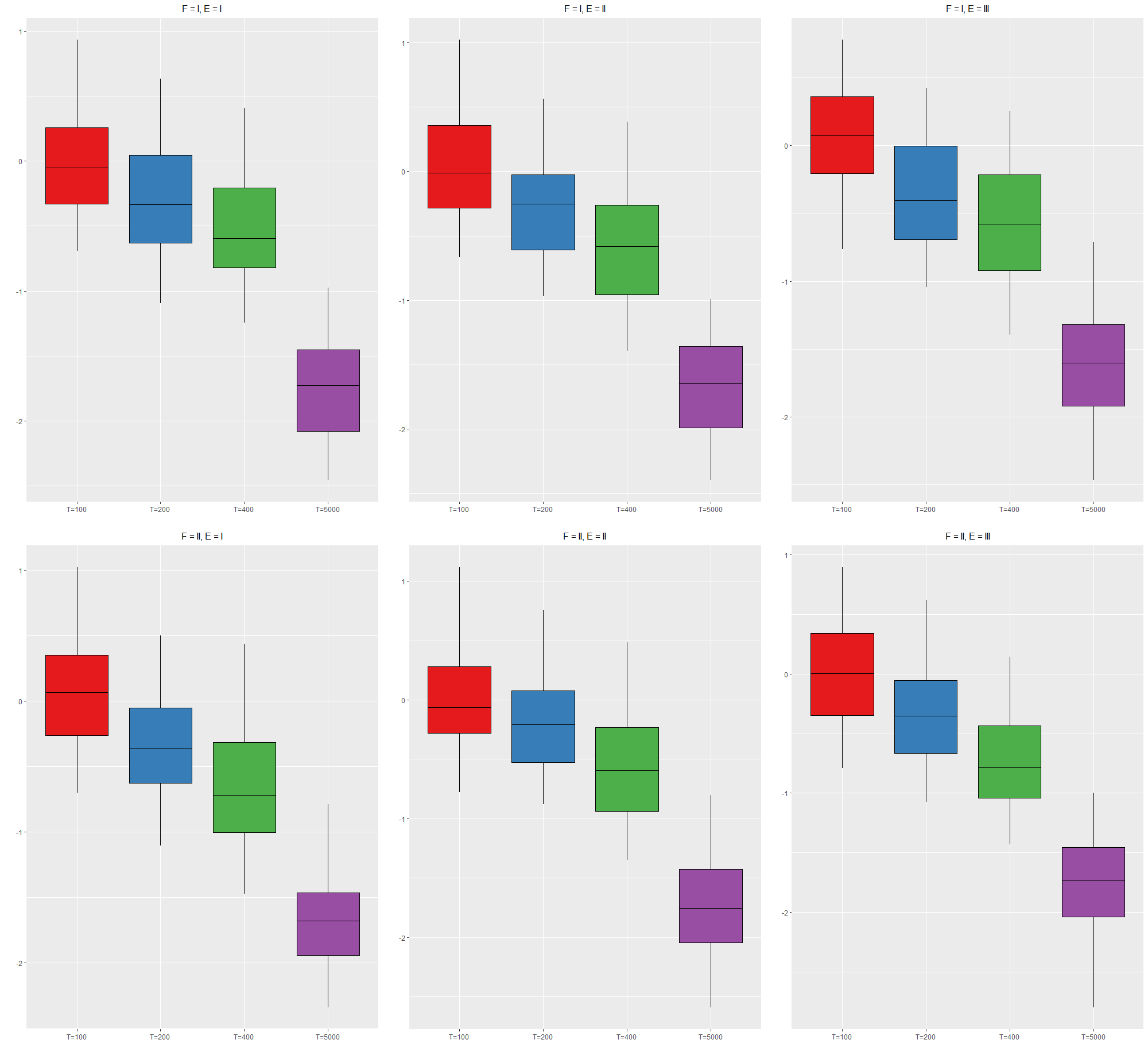}
    \caption{\setstretch{1.25}The boxplot of estimation loss for $\widehat{\bb} \otimes \widehat{\ba}$.}
    \label{Figure.2}
    \end{minipage}
\end{figure}

On the basis of the results from Table \ref{Table.1} and \ref{Table.2}, we next visualize the distribution of the estimation errors using boxplots Figure \ref{Figure.1} and \ref{Figure.2}. These plots offer a clearer view of the variability and central tendency of the errors, providing further insights into the robustness and consistency of our estimation procedure. The following part presents the boxplot of the estimation errors for $\wh\bb \otimes \wh\ba$ and \(\wh\bF_t\) under the same configurations as in Tables \ref{Table.1} and \ref{Table.2}.

In Figure \ref{Figure.1}, we present the boxplot of estimation errors for $\wh\bF_t$, following the same configurations as in Table \ref{Table.1}. Figure \ref{Figure.1} consists of 18 subplots, grouped in sets of three, each representing the estimation error boxplot for a specific configuration of $\bF_t$ and $\bE_t$, with $T = \{0.5pq, pq, 2pq\}$. Within each subplot, the three lines correspond to the settings $(p,q) = (5,10), (10,10), (20,20)$. From Figure \ref{Figure.1}, we can see that the estimation errors for $\wh\bF_t$ decrease as $(p,q)$ increases. This pattern holds across all subplots, further supporting the theoretical property of $\wh\bF_t$ discussed earlier.

In Figure \ref{Figure.2}, we present the boxplot of estimation errors for $\wh\bb \otimes \wh\ba$, based on the same simulation setup as in Table \ref{Table.2}. Figure \ref{Figure.2} consists of 6 subplots, each representing a distinct configuration of $\bF_t$ and $\bE_t$. Within each subplot, there are four lines, corresponding to $T = \{100, 200, 400, 5000\}$. For simplicity, we only show the results for $(p,q) = (20,20)$. From Figure \ref{Figure.2}, we observe that the estimation errors for $\wh\bb \otimes \wh\ba$ decrease as $T$ increases, a trend that holds across all subplots. This is consistent with the theoretical property of $\wh\bb \otimes \wh\ba$ discussed earlier.

\subsection{Asymptotic Normality}
\justifying
\setstretch{1.5}

Furthermore, we investigate the asymptotic normality of $\begin{pmatrix} \wh\ba \\ \wh\bb \end{pmatrix}$ and $\wh\bb \otimes \wh\ba$ under different values of $\alpha$ in the first step (\textit{$\alpha$-PCA}) of the estimation procedure. As an example, we use the settings $\bF_t \sim \text{i}$ and $\bE_t \sim \text{i}$, along with \((p, q)=(10,10)\), \((k,r) = (3,2)\) and time horizon \(T=400\). We conduct 1000 replications for each configuration. The \texttt{Q-Q} plots in Figure A.1 and histograms in Figure A.2 presented in Appendix demonstrate the asymptotic normality derived from Theorem \ref{T1}.


Figure A.1 presents two \texttt{Q-Q} plots for the first dimension of the estimated loading vectors, \(\wh\ba\), \(\wh\bb\) and \(\wh\bb \otimes \wh\ba\). In panel \textbf{(a)}, we report the \texttt{Q-Q} plots for the first dimension of \(\wh\ba\) and \(\wh\bb\) under three different values of hyperparameter \(\alpha = -1, 0, 1\), respectively. Each plot compares the ordered values of the respective estimates against the theoretical quantiles from a normal distribution, with the red line indicating the expected linear relationship under normality. From panel \textbf{(a)}, we observe that the empirical quantiles of the sample closely align with the theoretical quantiles of the normal distribution, indicating a strong adherence to normality. Panel \textbf{(b)} presents \texttt{Q-Q} plots for the first dimension of the Kronecker product \(\wh\bb \otimes \wh\ba\). From panel \textbf{(b)}, we observe that the \texttt{Q-Q} plots for this product also exhibit a similar pattern that in panel \textbf{(a)}. These findings are consistent with the asymptotic normality property of the estimators in Theorem \ref{T1}, providing evidence that for large sample sizes \(T\), the sample data behave asymptotically normal.


Figure A.2 presents histograms for the first dimension of the estimated loading vectors, \(\wh\ba\), \(\wh\bb\) and \(\wh\bb \otimes \wh\ba\). In panel \textbf{(a)}, we report the histograms for \(\wh\ba\), \(\wh\bb\) under three different values of hyperparameter \(\alpha = -1, 0, 1\), respectively. The histograms depict the distribution of the estimated values, with the black line representing the theoretical normal distribution for comparison. From Figure A.2, we observe that the empirical distribution of the sample data closely aligns with the theoretical probability density curve of the normal distribution, suggesting that the data approximates normality. Panel \textbf{(b)} presents histograms for the first dimension of the product \(\wh\bb \otimes \wh\ba\). From Figure A.2, we observe that these histograms follow a similar pattern to those in panel \textbf{(a)}. These findings are consistent with the asymptotic normality property of the estimators in Theorem \ref{T1}, providing additional evidence that, for large sample sizes \(T\), the estimators exhibit normal behavior.

In the subsequent subsection (Section \ref{sec5-4}), we validate the supervised screening refinement method introduced in Section \ref{sec4} and present the forecasting results using both the original and refined simulated data.

\subsection{Refinement with Supervised Screening}\label{sec5-4}
\justifying
\setstretch{1.5}

In the previous simulation estimation procedures, the estimation of \( \widehat{\bF}_t \) was based solely on the information contained in the observation matrix \( \bX_t \), without accounting for its potential correlation with the target variable \( y_t \). However, certain rows (countries) and columns (indicators) in \( \bX_t \) may exhibit weak correlations with \( y_t \), leading to redundancy in factor estimation and reduced forecasting efficiency. To address this, we introduced a supervised screening refinement procedure that incorporates the correlation structure between \( \bX_t \) and \( y_t \) as a filtering criterion, as detailed in Section \ref{sec4}.

To assess the effectiveness of this variable refinement process, we construct observation matrices containing partial noise and compare the performance of the estimation procedure with and without the screening step. We estimate the factor matrices from both the original observation matrix \( \bX_t \) and the refined matrix \( \widetilde{\bX_t} \), then evaluate the corresponding Mean Squared Forecasting Error (MSFE) for the dependent variable \( y_{t+h} \). The MSFE is calculated by
\begin{equation}\label{eqn9}
\displaystyle \text{MSFE} = \frac{1}{T-h}\sum_{t=1}^{T-h}({\wh y}_{t+h} - y_{t+h})^2.
\end{equation}

Without loss of generality, we consider an observation matrix with dimensions \( (p, q) = (10, 10) \). To introduce noise, we expand the matrix by adding 5 additional rows and 5 additional columns, where the entries are filled with independent and identically distributed noise drawn from the standard normal distribution \( \varepsilon \sim \mathcal{N}(0, 1) \). This results in a noisy observation matrix of dimension \( (15, 15) \). In other settings, the latent factor matrix \( \bF_t \in \mathbb{R}^{k \times r} \) is set with \( (k, r) = (3, 2) \), the forecasting step is defined with \( h = 1 \), and the total sample size is \( T = 1000 \). The hyperparameter \( \alpha \) for \textit{\( \alpha \)-PCA} is varied from \( -1 \) to \( 1 \) in increments of \( 0.1 \). We use the settings \( \bF_t \sim \text{ii} \) and \( \bE_t \sim \text{i}\). In the forecasting step, the sample is split into a training set and a test set in an 8:2 ratio. The latent factor matrix and loading vectors estimated from the training set are then applied to the test set for prediction, and the final MSFE is computed.

For the threshold parameter \( \rho_\delta \), we set \( \rho_\delta = 0.06 \) for both rows and columns, determined via cross-validation. Any row or column in the noisy matrix \( \bX_t \) is discarded if its average correlation with the target variable \( y_t \) is below \( \rho_\delta \). The simulation results presented in this section are based on 200 repetitions.

The comparison results for MSFE with noise matrix \(\bX_t\), refined matrix \(\widetilde\bX_t\), along with the percentage reduction in loss are presented in Table \ref{Table.3}.

\begin{table}[!ht]
  \centering
  \footnotesize
  \captionsetup{width=\textwidth}
  \caption{\setstretch{1.25}MSFE and percentage of loss reduction with noise matrix and refined matrix. We selected values for the hyperparameter \( \alpha \) in the range from -1 to 1 with an interval of 0.1. The minimum values of MSFE for both the noise matrix and refined matrix, as well as the maximum percentage of loss reduction, are highlighted in bold with a box. The threshold parameter \( \rho_\delta \) for correlation is set to 0.06. All computations are based on 200 replications.}
  \setstretch{1.25}
    \begin{tabular}{cccc}
    \toprule
    \textbf{$\alpha$} & MSFE with noise matrix & MSFE with refined matrix & Percentage of loss reduction\\
    \cmidrule{2-4}
    -1    & 15.712  & 8.415  & 46.4\% \\
    -0.9  & 10.510  & 7.299  & 30.6\% \\
    -0.8  & 10.871  & 6.907  & 36.5\% \\
    -0.7  & 14.006  & 8.627  & 38.4\% \\
    -0.6  & \fbox{\textbf{9.962}}   & 6.671  & 33.0\% \\
    -0.5  & 12.383  & 7.918  & 36.1\% \\
    -0.4  & 13.357  & 5.962  & 55.4\% \\
    -0.3  & 12.079  & 9.940  & 17.7\% \\
    -0.2  & 12.258  & 6.498  & 47.0\% \\
    -0.1  & 11.831  & 6.045  & 48.9\% \\
    0     & 18.168  & 7.141  & \fbox{\textbf{60.7\%}} \\
    0.1   & 13.503  & 6.115  & 54.7\% \\
    0.2   & 12.476  & 6.558  & 47.4\% \\
    0.3   & 14.408  & 7.626  & 47.1\% \\
    0.4   & 10.894  & \fbox{\textbf{5.776}}  & 47.0\% \\
    0.5   & 12.442  & 7.393  & 40.6\% \\
    0.6   & 12.965  & 6.111  & 52.9\% \\
    0.7   & 11.191  & 9.132  & 18.4\% \\
    0.8   & 13.982  & 9.292  & 33.5\% \\
    0.9   & 10.443  & 7.424  & 28.9\% \\
    1     & 14.081  & 6.090  & 56.8\% \\
    \bottomrule
    \end{tabular}%
  \label{Table.3}%
\end{table}%

In Table \ref{Table.3}, we report the MSFE values for both the noise matrix and the refined matrix after applying the supervised screening procedure, along with the associated percentage of loss reduction, under different configurations of \( \alpha \), which range from -1 to 1 with an increment of 0.1. From Table \ref{Table.3}, we observe that the minimum MSFE with the noise matrix occurs at \( \alpha = -0.6 \) with a value of 9.962, while the minimum MSFE with the refined matrix is observed at \( \alpha = -0.4 \), yielding a value of 5.776. The highest percentage of loss reduction, 60.7\%, is achieved at \( \alpha = 0 \). Across all configurations, the MSFE demonstrates varying degrees of improvement in predictive accuracy. The most significant enhancement occurs when \( \alpha = 0 \), with a reduction of 60.7\%, which substantially improves the forecast accuracy of \( y_{t+h} \). These simulation results confirm the effectiveness of our proposed method, which successfully eliminates less relevant components from the observation matrix, leading to a considerable reduction in prediction error and improved forecasting accuracy for \( y_{t+h} \).

\section{Empirical Analysis}\label{sec6}

\subsection{Data}
\justifying
\setstretch{1.5}

In this section, we evaluate the empirical effectiveness of the proposed model using real-world data. The dataset, sourced from the OECD, includes 10 quarterly macroeconomic indicators for 14 countries, spanning from 1993.Q1 to 2019.Q3. This results in a matrix-valued time series with dimensions \(T = 107\) and \(p \times q = 14 \times 10\), analogous to the dataset employed in \cite{chen2023statistical}. The 14 countries included in the analysis are the United States (USA), Canada (CAN), New Zealand (NLD), Australia (AUS), Norway (NOR), Ireland (IRL), Denmark (DNK), the United Kingdom (GBR), Finland (FIN), Sweden (SWE), France (FRA), the Netherlands (NZL), Austria (AUT), and Germany (DEU). The 10 indicators are organized into four categories: production (P: TIEC, P: TM, GDP), consumer prices (CPI: Food, CPI: Ener, CPI: Tot), money market (IR: Long, IR: 3-Mon), and international trade (IT: Ex, IT: Im). To meet the mixing conditions outlined in Assumption \ref{A1}, each original univariate time series is transformed through first or second differencing or logarithmic transformation, followed by centralization to ensure stability in forecasting. The target variable for prediction is the quarterly real GDP growth of OECD countries (OECD: GDPG). Detailed descriptions and transformations of the dataset can be found in the Appendix.

\subsection{Results of Original Observations}
\justifying
\setstretch{1.5}

The sample is divided into a training set and a testing set in an 8:2 ratio, resulting in sample size with \(\{\text{Train}, \text{Test}\} = \{86, 21\}\). To assess prediction performance, we compute the Mean Squared Forecast Error (MSFE) as defined in Equation (\ref{eqn9}) on the testing set under seven different configurations of factor dimensions \((k, r)\), along with the factor matrix estimated as described in Section \ref{subsec2.3} marked with a superscript \(^*\) in Table \ref{Table.4}, where \((k, r)\) indicates the row and column dimensions of the estimated factor matrix.

For comparison, we consider four benchmark models. (1) The original data model \textit{(Raw)}. In this model, \(y_{t+h}\) is forecasted using only the original observation matrix \(\bX_t\), represented as \(y_{t+h} = \ba^{\prime} \bX_t \bb + \varepsilon_{t+h}\), without any transformation. (2) The vectorized data model \textit{(Vec)}. Here, \(y_{t+h}\) is forecasted using the vectorized form of the original observation matrix, \(\mathrm{vec}(\bX_t)\), such that \(y_{t+h} = \bb^{\prime} \mathrm{vec}(\bX_t) + \varepsilon_{t+h}\). (3) The vectorized data model with Lasso \textit{(Vec-Lasso)}. This model is similar to (2), but the coefficients are estimated using the Lasso method in \cite{tibshirani1996regression}, so \(y_{t+h} = \bb_{l}^{\prime} \mathrm{vec}(\bX_t) + \varepsilon_{t+h}\), where \(\bb_{l}\) denotes the Lasso-estimated coefficients for \(\bb\) from model (2). (4) The autoregression model \textit{(AR)}. In this model, \(y_{t+h}\) is forecasted based on lagged values of \(y_t\). For simplicity, we use the AR(1) model as a baseline, where \(y_{t+1} = \alpha y_t + \varepsilon_{t+1}\), and \(\alpha\) represents the autoregressive coefficient. The train-test split proportions used in these models are the same as those in our proposed model. We refer to our model as \(\alpha\)-\textit{PCA-LSE}.

The MSFE comparison across different models and configurations are reported in Table \ref{Table.4}.

\begin{table}[!ht]
\centering
\footnotesize
\captionsetup{width=\textwidth}
\caption{\setstretch{1.25}MSFE with original observations across various configurations and models. Panel A presents the forecasting results of our model \(\alpha\)-PCA-LSE for different values of \(\alpha\) and factor dimensions \((k, r)\), with the estimated \((k, r) = (2,2)\) denoted by a superscript \(^*\). Panel B shows the MSFE results for various benchmark models, including the original data model (Raw), the vectorized data model (Vec), the vectorized data model with Lasso (Vec-Lasso), and the autoregression model (AR). These benchmark models do not include the hyperparameter \(\alpha\) or latent factors \((k, r)\) in their forecasting procedure, and therefore, the MSFE is reported as a single value for each model. All reported values are the averages over the 10-fold CV with rolling window.}
\setstretch{1.25}
\begin{tabular}{cccccccccc}
\toprule
\multicolumn{10}{l}{Panel A: \(\alpha\)-PCA-LSE} \\
\midrule
\multirow{6}{*}{\(\alpha\)-PCA-LSE} & \(\alpha\) & (6,5) & (5,5) & (4,5) & (4,4) & (3,4) & (3,3) & (3,2) & *(2,2) \\
\cmidrule{2-10}
& -1  & 0.167  & 0.169  & 0.397  & 0.316  & 0.601  & 0.792  & 0.258  & 0.263  \\
& -0.5  & \fbox{\textbf{0.166}}  & 0.166  & 0.380  & 0.287  & 0.577  & 0.735  & 0.257  & 0.265  \\
& 0 & 0.299  & 0.286  & 0.400  & 0.257  & 0.557  & 0.680  & 0.255  & 0.269  \\
& 0.5 & 0.363  & 0.447  & 0.661  & 0.890  & 0.541  & 0.629  & 0.254  & 0.272  \\
& 1   & 0.208  & 0.189  & 0.751  & 0.219  & 0.531  & 0.582  & 0.254  & 0.272  \\
\midrule
\multicolumn{10}{l}{Panel B: Benchmarks} \\
\midrule
& & & Raw & Vec & Vec-Lasso & AR \\
\cmidrule{4-7}
& & MSFE & 3.480 & 698.924 & 14.786 & 1.005 \\
\bottomrule
\end{tabular}
\label{Table.4}
\end{table}

In Table \ref{Table.4}, we report the forecasting results, measured by MSFE, for our proposed model \(\alpha\)\textit{-PCA-LSE} across different configurations, alongside four benchmark models, with original observations. These benchmark models include the original data model (\textit{Raw}), the vectorized data model (\textit{Vec}), the vectorized data model with Lasso regularization (\textit{Vec-Lasso}), and the autoregressive model (\textit{AR}). For simplicity, we select hyperparameters \(\alpha\) in the range from \(-1\) to \(1\) with intervals of 0.5, and we use seven different fixed dimensions for the latent factor matrix, given by \((k,r)=\{(6,5), (5,5), (4,5), (4,4), (3,4), (3,3), (3,2)\}\). Furthermore, the estimated dimensions for the latent factor matrix, based on Section \ref{subsec2.3}, are \((k,r) = (2,2)\), denoted by a superscript \(^*\). The penalty coefficient in the \textit{Vec-Lasso} model is selected via cross-validation. From Table \ref{Table.4}, we observe that the configuration \((k,r)=(6,5)\) with \(\alpha = -0.5\) minimizes the overall forecasting loss (MSFE) among all models, achieving a value of 0.166, as indicated by the bold text in the box. The results for the benchmark models are relatively large compared to our proposed model, especially for the \textit{Vec} model, which yields a MSFE of 698.924. This is due to the high dimension of the vectorized matrix, which leads to sparsity issues during the estimation process with the training data. In contrast, the results are significantly improved in the \textit{Vec-Lasso} model with the regularization procedure applied during the Lasso estimation. The Diebold-Mariano test (\cite{diebold1991comparing}) shows that the p-values for the comparisons between our model and the benchmark models are 0.033, 0.008, 0.022, and 0.082 for \textit{Raw}, \textit{Vec}, \textit{Vec-Lasso}, and \textit{AR}, respectively, all significant at the 10\% level, indicating that our forecasting method outperforms the benchmarks. These forecasting results are consistent with our previous analysis of the dimension reduction procedure, further demonstrating that our model outperforms the benchmark models. 

\subsection{Results of Supervised Screening Refinement Observations}
\justifying
\setstretch{1.5}

\begin{figure}[!ht]
    \centering
    \includegraphics[width=0.9\linewidth]{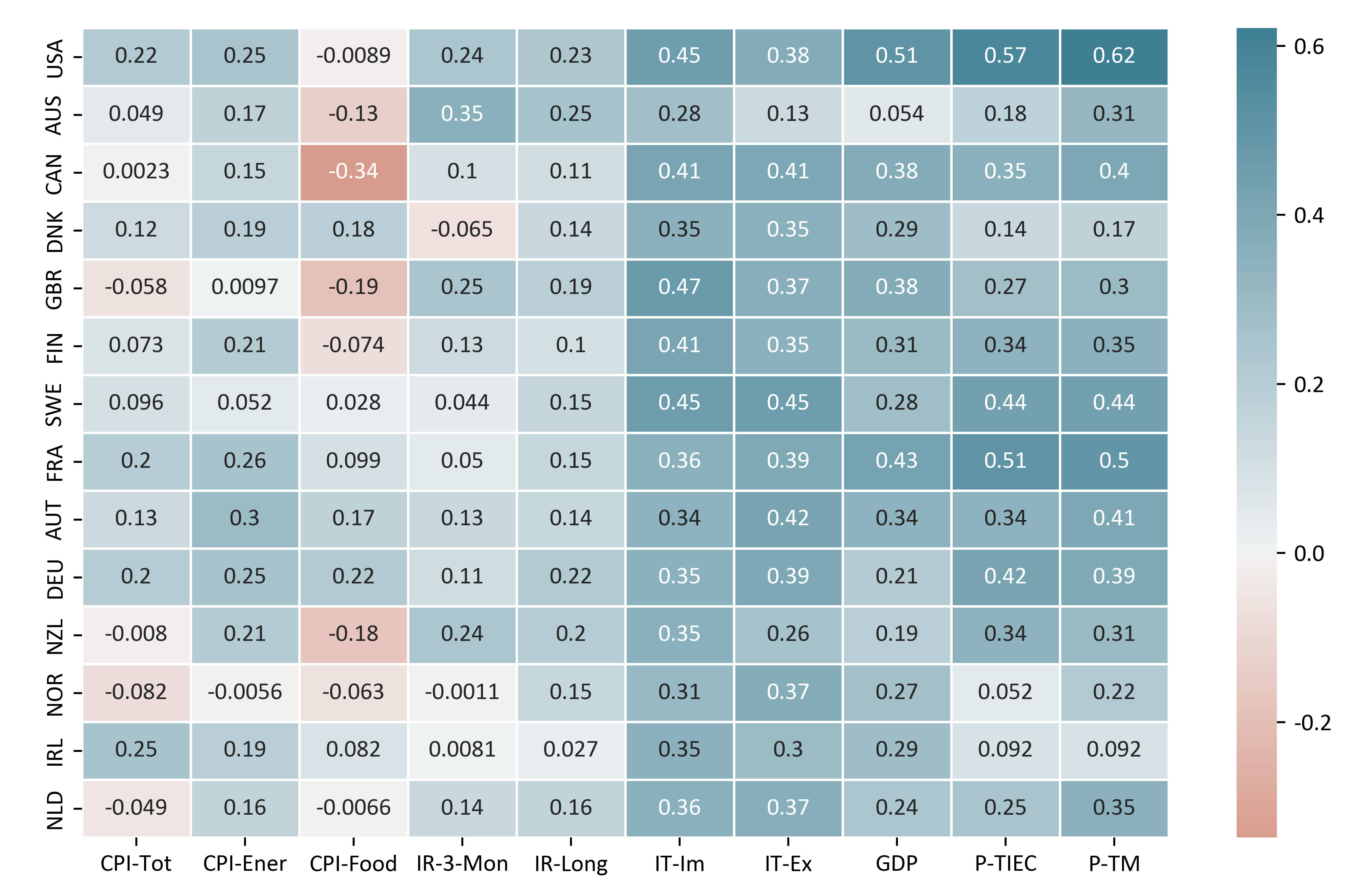}
    \caption{\setstretch{1.25}Heatmap of correlation coefficients between target indicators and elements of the original observation matrix across different countries and economic variables. The rows of the correlation matrix represent the 14 different countries, while the columns correspond to the 10 distinct economic indicators.}
    \label{Figure.5}
\end{figure}

In Section \ref{sec4}, we discuss the potential benefits of the supervised screening procedure, which utilizes the correlations between the target indicator and the elements of the observation matrix. In this section, we apply this screening approach to our empirical data analysis. Specifically, we first calculate the pairwise correlations, which are illustrated in Figure \ref{Figure.5} as follows.

In Figure \ref{Figure.5}, we calculate the correlation coefficients between the target indicators and the elements of the original observation matrix across different countries and economic variables, denoted as \(\bar\rho_i\) and \(\bar\rho_j\) in Section \ref{sec4}. From Figure \ref{Figure.5}, we observe that the correlation between certain pairs, such as CPI: Food in Canada and IR: 3-Mon in Denmark, shows a significant negative value, while other pairs, such as GDP and P: TM across all countries, exhibit positive correlations with the target indicators. Furthermore, Figure \ref{Figure.5} reveals some rows and columns with low correlation coefficients, such as those for Ireland (IRL) and IR: 3-Mon. This suggests that rows and columns with low correlation coefficients may be excluded, retaining only the most informative variables in the observation matrix structure, which could potentially enhance the accuracy of the forecasting procedure.

To determine which rows and columns should be retained, we calculate the average correlation coefficients across 14 rows representing countries and 10 columns representing indicators. These results are then presented in two line chart subfigures, labeled as \texttt{Country} and \texttt{Index} on the x-axis, respectively. Figure A.3 provides a detailed representation of these outcomes.


In Figure A.3, we present the average correlation coefficients across the rows (\texttt{Country}) and columns (\texttt{Index}) in descending order, denoted as \(\bar\rho_i\) and \(\bar\rho_j\) in Section \ref{sec4}, respectively. From Figure A.3, we observe that the indicators, such as IR: IM, IT: EX, and P: TM, show the highest correlation values, approximately 0.35, indicating stronger average correlations. Moreover, we note that the correlation value for GDP drops sharply from 0.30 to 0.17 and stabilizes with the next two indicators. Consequently, we exclude the columns corresponding to the indicators following IR: Long. In parallel, the USA exhibits the highest correlation with the target variable, while countries such as Australia (AUS) display significantly lower correlation values. This suggests that countries with correlations lower than those of Australia can be excluded from the analysis. To achieve this, we remove the rows and columns with low correlation, specifically where \({\bar{\rho}}_i\) (\({\bar{\rho}}_j\)) fall below thresholds of \({\bar{\rho}}_i < 0.19\) and \({\bar{\rho}}_j < 0.16\). After this screening process, we retain 12 countries (USA, FRA, DEU, AUT, CAN, GBR, SWE, FIN, NZL, NLD, DNK, AUS) in the rows and 7 indicators (IT: IM, IT: EX, P: TM, P: TIEC, GDP, CPI: Enter, IR: Long) in the columns. Thus, the resulting screened matrix \(\widetilde{\bX}_t\) is of size \(\mathbb{R}^{\widetilde{p} \times \widetilde{q}}\), where \((\widetilde{p}, \widetilde{q}) = (12, 7)\).

Next, we present our forecasting results with supervised screening refined observation matrix in Table \ref{Table.5}.

\begin{table}[!ht]
\centering
\footnotesize
\captionsetup{width=\textwidth}
\caption{\setstretch{1.25}MSFE with supervised screening refinement observation matrix across various configurations and models. The configurations in Panel A are identical to those presented in Table \ref{Table.4}, and therefore, no further details will be provided here. Panel B presents the results for the benchmark models, where the MSFE for four comparison models is computed, as shown in Table \ref{Table.4}. The transformation of the observation matrix does not impact the forecasting procedure of the AR model, and thus, the result remain consistent with that in Table \ref{Table.4}. All reported values are the averages over the 10-fold CV with rolling window.}
\setstretch{1.25}
\begin{tabular}{cccccccccc}
\toprule
\multicolumn{10}{l}{Panel A: \(\alpha\)-PCA-LSE} \\
\midrule
\multirow{6}{*}{\(\alpha\)-PCA-LSE} & \(\alpha\) & (6,5) & (5,5) & (4,5) & (4,4) & (3,4) & (3,3) & (3,2) & *(2,2) \\
\cmidrule{2-10}
& -1    & 0.330  & 0.297  & 0.278  & 0.403  & 0.214  & 0.176  & 0.180  & 0.166  \\
& -0.5  & \fbox{\textbf{0.138}}  & 0.319  & 0.436  & 0.297  & 0.153  & 0.170  & 0.186  & 0.169  \\
& 0     & 0.195  & 0.277  & 0.869  & 0.198  & 0.166  & 0.181  & 0.174  & 0.176  \\
& 0.5   & 0.189  & 0.276  & 0.847  & 0.197  & 0.164  & 0.181  & 0.172  & 0.175  \\
& 1     & 0.225  & 0.269  & 0.230  & 0.177  & 0.154  & 0.169  & 0.196  & 0.173  \\
\midrule
\multicolumn{10}{l}{Panel B: Benchmarks} \\
\midrule
& & & Raw & Vec & Vec-Lasso & AR \\
\cmidrule{4-7}
& & MSFE & 2.476 & 18.146 & 2.607 & 1.005 \\
\bottomrule
\end{tabular}
\label{Table.5}
\end{table}

In Table \ref{Table.5}, we present the forecasting performance, quantified by the Mean Squared Forecast Error (MSFE), for the proposed \(\alpha\)\textit{-PCA-LSE} model under various parameter settings, together with four benchmark models, all using the screened observation matrix. From Table \ref{Table.5}, we see that the forecasting loss is substantially and uniformly reduced for all models except the \textit{AR} model, which does not include the observation matrix in its forecasting process. The minimum MSFE is attained by our proposed \(\alpha\)\textit{-PCA-LSE} model with \(\alpha = -0.5\) and factor dimensions \((k,r) = (6,5)\), which are consistent with the configuration reported in Table \ref{Table.4}. Under this setting, the MSFE reaches a minimum value of 0.138, marked in bold in the box, demonstrating that our model outperforms the alternative approaches. The estimated latent factor dimensions remain \((k,r) = (2,2)\), indicated by the superscript \(^*\). 

The empirical evidence suggests that the supervised screening procedure effectively reduces the estimation error (MSFE), highlighting the benefit of incorporating target-variable information to remove low-correlation components from the original observation matrix. This enables the model to retain only the most informative rows and columns, resulting in improved forecasting precision. The Diebold-Mariano test shows that the p-values for the benchmark models compared to ours are 0.040, 0.000, 0.000, and 0.017 for \textit{Raw}, \textit{Vec}, \textit{Vec-Lasso}, and \textit{AR}, respectively, all significant at the 5\% level. This observation supports our previous theoretical discussion, showing that, compared to the standard unsupervised PCA, the supervised screening method exploits the relationship between predictors and the target variable to achieve targeted dimension reduction. This in turn reduces the complexity of models and significantly enhances its predictive performance. In this study, we construct the screening threshold using a Scree plot-based approach. For future research, it would be worthwhile to develop more advanced techniques for determining the refined matrix and its optimal threshold to further strengthen the effectiveness of the screening process and dimension reduction.

\section{Conclusion and Discussion}\label{sec7}
\justifying
\setstretch{1.5}

Building on previous research, this study extends the diffusion index model to the setting of high-dimensional matrix-variate time series data. To this end, we propose a new methodology that utilizes the \(\alpha\)-PCA technique for dimension reduction of the observation matrix, thereby extracting a low-rank factor structure. Subsequently, an iterative least-squares estimation (LSE) approach is adopted to estimate the model parameters, allowing us to recover the latent factor process and achieve more effective indicator forecasting.

In addition, this paper puts forward a novel supervised screening refinement strategy for the observation matrix. By applying this procedure, we systematically compare the prediction losses obtained using the original matrix versus those obtained using the refined matrix. Simulation evidence shows that the finite-sample performance aligns well with the corresponding theoretical properties, which further supports the asymptotic validity of the proposed estimator.

The empirical analysis demonstrates that the matrix prediction model incorporating factor dimension reduction outperforms conventional benchmark models. Importantly, the prediction accuracy is further enhanced when the supervised screening procedure is employed, suggesting that removing low-correlation rows and columns effectively improves forecasting precision. This highlights the critical role of matrix refinement in mitigating noise and isolating informative dimensions, thereby increasing the reliability of forecast results.

As directions for future research, extending this framework to incorporate higher-frequency data, such as daily or intra-day series, would enable finer-grained prediction. Additionally, integrating advanced deep learning techniques within the matrix-factor structure could capture more complex, nonlinear relationships between the observation matrix and the target variable, potentially improving predictive performance and robustness.





\section*{Conflict of Interest Statement}
We declare that there are no relevant financial or non-financial competing interests to disclose.


\section*{Data Availability Statement} 
The data supporting the findings of this study are openly accessible at  OECD Data Explorer (\url{https://data-explorer.oecd.org/vis?tenant=archive&df[ds]=DisseminateArchiveDMZ&df[id]=DF_EPER&df[ag]=OECD}).



\onehalfspacing
\footnotesize
\bibliographystyle{econometrica}
\bibliography{reference}
\onehalfspacing

\end{document}